\documentstyle[emulateapj,rotating]{article}
 
\begin{document}
 
\lefthead{\hspace{-0.2cm} No. 1, 2000 \hspace{3.25cm} COSMIC CONCORDANCE AND QUINTESSENCE  \hspace{5.0cm} {}}
\righthead{\hspace{7.75cm} WANG ET AL. \hspace{6.5cm} Vol. 530}

\submitted{Received 1999 January 27; accepted 1999 September 29}
 
\title{ Cosmic Concordance and Quintessence }

\author{Limin Wang\altaffilmark{1}, 
	R. R. Caldwell\altaffilmark{2},
	J. P. Ostriker\altaffilmark{3},
	and Paul J. Steinhardt\altaffilmark{2}}

\altaffiltext{1}{Department of Physics,
		Columbia University,
		538 West 120th Street, New York, NY 10027} 
\altaffiltext{2}{Department of Physics,
		Princeton University,
		Princeton, NJ 08544}
\altaffiltext{3}{Department of Astrophysical Sciences,
		Princeton University,
		Princeton, NJ 08544}

\begin{abstract}

We present a comprehensive study of the observational constraints on spatially
flat  cosmological models containing a mixture of matter and  quintessence ---
a time varying, spatially inhomogeneous component of the energy density of the
universe with negative pressure.  Our study also includes the limiting case of
a  cosmological constant.   We classify the observational constraints by red
shift:  low red shift constraints include the Hubble parameter, baryon
fraction, cluster abundance,  the age of the universe, bulk velocity and the
shape of the mass power spectrum; intermediate red shift constraints are due to
probes of the red shift-luminosity distance based on type 1a supernovae, 
gravitational lensing, the Lyman-alpha forest, and the evolution of large scale
structure; high red shift constraints are based on measurements of the cosmic
microwave background temperature anisotropy.  Mindful of systematic errors, we
adopt a conservative approach in applying these observational constraints.  We
determine that the range of quintessence models in which the ratio of the
matter density to the critical  density is  $0.2 \lesssim \Omega_m \lesssim
0.5$ and the effective, density-averaged equation-of-state is $-1 \le w
\lesssim -0.2$, are consistent with the most reliable, current low red shift
and microwave background observations at the $2 \sigma$ level. Factoring in the
constraint due to the recent measurements of type 1a supernovae, the range for
the equation-of-state is reduced to $-1 \le w \lesssim -0.4$, where this range
represents models consistent with each observational constraint at the
2$\sigma$ level or better (concordance analysis). A combined maximum
likelihood  analysis suggests a smaller range, $-1 \le w \lesssim -0.6$.   We
find that the best-fit and best-motivated  quintessence models lie  near
$\Omega_m \approx 0.33$, $h \approx 0.65$, and spectral index $n_s=1$, with an
effective  equation-of-state $w \approx -0.65$ for ``tracker'' quintessence and
$w=-1$ for ``creeper'' quintessence.

\end{abstract}
  
\section{Introduction}

The most widely studied cosmological models at the present time are variants of
the Cold Dark Matter (CDM) paradigm within which adiabatic perturbations in a
dominant CDM species grow due to gravitational instability from quantum
fluctuations imprinted during an inflationary era.  The bulk of the evidence
today strongly favors models within which $\Omega_m <1$ and any hot component
is not significant, $\Omega_{\rm HDM} \ll \Omega_{\rm CDM}$. Several authors
(\cite{OS95,KT95,TurnerWhite97}) have found that the best and simplest fit
concordant with current observations is provided by
\begin{eqnarray}
\Omega_m &=& [\Omega_{CDM}] + [\Omega_{baryon}] 
\cr\cr &\approx& [0.30 \pm 0.10] + [0.04 \pm 0.01]
\end{eqnarray}
with $h=0.65 \pm 0.15$.  One is thus led to either an open universe or one in
which the remaining energy density required to produce a geometrically flat
universe is some additional energy component ($E$) with
\begin{equation}
\Omega_m + \Omega_E = 1.
\end{equation}
An important advantage of the flat model is that it is consistent with standard
inflationary cosmology, and its associated resolution of the cosmological
horizon and flatness problem and prediction of a nearly scale-invariant
spectrum of energy density fluctuations. Finding these arguments
compelling\footnote{ While inflationary cosmologies  can be constructed which
lead to a spatially open universe  (\cite{Gott82}, and {\it e.g.}
\cite{LindeOpen} and references therein),  provided careful tuning, choice of
inflaton potential, and/or anthropic arguments, we consider here only the
standard case of a spatially flat universe.},  we will adopt the ansatz that
the universe is spatially flat.
  
In our previous work (\cite{OS95}) we identified the range of models consistent
with then-current observations, restricting attention to the case where
$\Omega_E$ is vacuum energy or cosmological constant ($\Lambda$).  The
cosmological  constant, a static, homogeneous energy component with positive
energy density but negative pressure, was introduced initially by Einstein in a
flawed attempt to model our universe as a static spacetime with positive
spatial curvature. In a spatially flat universe, however, the negative
pressure  results in a repulsive gravitational effect which accelerates the
expansion of the universe.    This earlier analysis, which preceded by several
years the recent  evidence based on luminosity-red shift surveys  of type 1a
supernovae,  already found that $\Lambda$ is favored over standard cold dark
matter or open models.  Over the intervening years, the supernovae 
measurements  as well as  other observations have reinforced these conclusions.

In this paper, we update our earlier concordance analysis and expand it to
include the possibility that the additional energy component consists of 
quintessence, a dynamical, spatially inhomogeneous  form of energy with
negative pressure (\cite{CDS98}). A common example is the energy of a slowly
evolving scalar field with positive potential energy,  similar to the inflaton
field in inflationary  cosmology
(\cite{Weiss87,RatraPeebles88,Wetterich95,FriemanEtAl95,CobleEtAl97,FerreiraJoyce97,CDS98,ZlatevEtAl98}).
Unlike a cosmological constant, the dynamical field can support long-wavelength
fluctuations which leave an imprint on the cosmic microwave background (CMB)
and large-scale structure. In particular, the long wavelength fluctuations
change the relation between the amplitude of the CMB anisotropy and the
gravitational potential fluctuations so that the COBE normalization of the mass
power spectrum depends on the pressure of the quintessence component. A further
distinction is that $w$, the ratio of pressure ($p$) to energy density
($\rho$), is $-1<w \le 0$ for quintessence whereas $w$ is precisely $-1$ for a
cosmological constant. Hence, the expansion history of the universe for a
$\Lambda$ model versus a quintessence model (for the same $\Omega_m$ today,
say) is different. In general, the acceleration, the age and the volume of the
universe are less for quintessence models than for $\Lambda$ models (assuming
all other cosmic parameters are fixed).

A prime motivation for considering quintessence models is to address the
``coincidence problem," the issue of explaining the initial conditions
necessary to yield the near-coincidence of the densities of matter and the
quintessence  component today. For the case of a cosmological constant, the
only possible option is to finely tune the  ratio of vacuum density to
matter-radiation density to $1$ part in $\sim 10^{120}$ at the close of
inflation in order to have the correct ratio today. Symmetry arguments from 
particle physics are sometimes invoked to explain why the cosmological constant
should be zero (\cite{Banks96}) but there is no known explanation for a
positive, observable vacuum density. For quintessence, because it couples
directly to other forms of  energy, one can envisage the possibility of
interactions which may cause the quintessence component to naturally adjust
itself to be comparable to the matter density today. In fact, recent
investigations (\cite{ZlatevEtAl98,SteinhardtEtAl98}) have introduced the
notion of ``tracker field" models of quintessence which have attractor-like
solutions  (\cite{PeeblesRatra88,RatraPeebles88}) which produce the current
quintessence energy density without the fine-tuning of initial conditions. A
related development has been ``creeper field'' models (\cite{HueyPJS}) which
are nearly as insensitive to initial conditions but indistinguishable from
$\Lambda$ today.

Fundamental physics provides some further motivation for light scalar fields. 
Particle physics theories with dynamical symmetry breaking or non-perturbative
effects have been found which generate potentials with ultra-light masses which
support negative pressure
(\cite{Affleck85,HillRoss88a,HillRoss88b,Binetruy96,Barreiro98,Binetruy98}).
These suggestive results lend appeal to a particle physics basis for
quintessence, as a logical alternative to an {\it ad hoc} invocation of a
cosmological constant.  We do not aim to base our investigation of the
properties of quintessence cosmologies on a specific particle physics model,
however, as such models are still in a developmental stage.  An intriguing 
thought is that progress in the  cosmological observations and experiments
discussed here will soon decide the issue, possibly pointing to new fundamental
physics inaccessible in the accelerator laboratory. We would emphasize that
scalar field models of quintessence are not only the simplest, well-motivated
choice  from a particle physics standpoint, but, also, they can mimic fluids
with arbitrary equation-of-state.  (It was shown explicitly in \cite{CDS98} 
that there is a one-to-one correspondence between a general time-dependent
equation-of-state  and an equivalent scalar field potential.)  Generalizations
to tensor fields or more general stress tensor (\cite{Hu99}) or topological
defects (\cite{Spergel96}) have also been considered; for the  purposes of this
study based on current observations, they can be well described by scalar
fields, in addition.

Extending the realm of cosmological models to include quintessence opens up a
new degree of freedom, the equation-of-state of the missing energy component.
The added degree of freedom  necessarily makes the procedure of selecting
viable models  more complicated.  Previous studies have considered some
specific  combinations of observations
(\cite{SilveiraWaga97,TurnerWhite97,Garnavich98,Perlmutter98}) or some specific
models (\cite{FerreiraJoyce97,FerreiraJoyce98}). Here we systematically examine
the most complete range yet of measures of astrophysical phenomena at low,
intermediate, and high red shift and make a complete search in parameter space
to objectively identify the viable models.  We show that the observations are
consistent with $\Lambda$ and quintessence models for a substantial range of
parameters. An impressive feature is how a number of observations, not only the
measurements of type 1a supernovae,  favor a missing energy component with
substantially negative pressure.

The pace of cosmological observations is proceeding so rapidly that any
quantitative conclusions may soon become dated. Nevertheless, we think that an
assessment at the present time is worthwhile for at least three reasons. First,
our study shows a concordance among a growing number of  observations. Compared
to the previous analysis (\cite{OS95}) new constraints have been added and old
ones have been revised, and, yet, notably, the key conclusions  ---  a new
energy component and an accelerating expansion rate --- have been significantly
strengthened. Second, the study isolates those observations which are playing
the lead roles in shaping the current conclusions and identifies   observations
or combinations thereof which will be most decisive in the near future. Third,
this comprehensive analysis enables one to identify specific best-fit models
which can be explored in much greater detail to search for more subtle
implications and tests.

The organization of the paper is as follows. In section \ref{parameters}, we
discuss the parameterization of the cosmological constant and quintessence
cosmological models. In section \ref{constraintsection}, we present the
observational constraints, classified by low, intermediate, and high red shift.
We evaluate the constraints, presenting the results in section
\ref{resultsection}. We conclude in section \ref{discussion} with an
identification of the overall best-fit models and a discussion of future
observations.

\section{Parameterization of Quintessence Cosmological Models}
\label{parameters}

The quintessence (QCDM) cosmological scenario is a spatially-flat FRW
space-time dominated by radiation at early times, and cold dark matter (CDM)
and quintessence (Q) at late times. For simplicity, we will consider models in
which the quintessence component consists of a scalar field slowly rolling down
its effective potential with a constant equation-of-state.  The detailed
equations-of-motion are discussed in \cite{CDS98}.  This class of models is a
good approximation for most of the range of  quintessence candidates. The
models we consider can then be fully characterized by the following five
parameters:

${\bf w}_{\bf Q}$:  A constant quintessence equation-of-state, in the range $w
\in [-1,0]$ in the present epoch.  In most cases, the quintessence
equation-of-state changes slowly with time,  but the observational predictions
are well approximated by treating $w$ as a constant, equal to
\begin{equation} 
\widetilde w \approx \int da \, \Omega_Q(a) \, w(a) / \int da \, \Omega_Q(a).
\label{effect}
\end{equation}
We will comment on exceptions later. Tracker field quintessence models have a
lower bound on the range of $w$, while creeper quintessence leads to $w$ very
close to $-1$, which is effectively indistinguishable from  a true cosmological
constant.

${\bf\Omega}_{\bf m}$:  The matter density parameter, defined as the ratio of
the matter energy density, including CDM and baryons, to the critical energy
density $\rho_c=3H^2/8\pi G$, where $H$ is the Hubble constant.  We assume
that  any contribution to the energy density due to a hot component, such as
neutrinos, is small, insofar as free-streaming has a negligible effect on the
clustering of CDM. Unless otherwise specified, we have imposed $\Omega_m +
\Omega_Q =1$, where $\Omega_Q$ is the corresponding density parameter for
quintessence.  Hence, the matter density parameter lies in the range $\Omega_m
\in [\Omega_b, 1]$.

${\bf \Omega}_{\bf b}$:  The baryon density parameter, defined as the ratio of
the baryonic energy density to the critical energy density.

${\bf h}$:  The Hubble parameter, related to the Hubble constant by $H=100\,
h$~km/s/Mpc.

${\bf n}_{\bf s}$:  The index of the power spectrum of primordial density
fluctuations in the matter and radiation. This parameter also controls the
contribution of tensor perturbations, for which we consider two cases.  In the
first case, we impose the inflationary relation between the amplitude of the
primordial density and gravitational wave (tensor) perturbations for $n_s \le
1$, revised for the case of quintessence (\cite{CS}).  For $n_s > 1$, we assume
the tensor contribution is negligible.  In the second case, we assume the
tensor contribution is negligible for all values of the spectral index.  (We
only illustrate the first case; the  second case yields an indistinguishable
result in regard to our concordance analysis.)

The most revealing way to depict the concordance of constraints on quintessence
models is to project the five-dimensional parameter space into the
$\Omega_m$-$w$ plane. In displaying this plane, we assume the universe is
spatially flat. Since the flatness condition requires $\Omega_Q = 1 -
\Omega_m$, the parameters $w$ and $\Omega_m$ completely specify the
quintessence portion of the cosmology.  $\Lambda$CDM corresponds to the line
$w=-1$, and SCDM corresponds to the line $\Omega_m=1$ in this plane.

\section{Observational Constraints}
\label{constraintsection}

We take a conservative approach in applying the cosmological constraints. 
Observational cosmology is currently in a period of rapid growth so that the
current constraints must be considered as {\it work in progress}, rather than
final.  Certainly, measurements of many astrophysical phenomena are becoming
more refined, with greater precision as statistical errors are reduced, and
with greater accuracy as systematic errors are better understood.  If this
situation described all observations, we could confidently apply the results to
the full limits of the published errors.  Because the observational constraints
typically restrict combinations of independent model parameters, then by
combining several experimental results, we could find tighter parameter bounds
than if the observations were applied individually.  For example, combining two
constraints restricts a two-parameter system to an ellipse in parameter space,
whereas the individual constraints applied successively allow a rectangle which
contains that ellipse.  The former combination would allow the determination of
a model which was ``best'' in the maximum likelihood sense.  However, not all
measurements have well controlled systematic errors or even high precision
statistical errors.  Nor do they have true gaussian errors. Nor are the errors
uncorrelated. A prime example is the current set of type 1a supernovae
magnitude-red shift data, which must be interpreted with caution.  For these
reasons, we do not advocate cosmological parameter extraction with the current
set of astrophysical data. Until the coming generation of precision experiments
are on-line, we believe it can be misleading to combine the full set of
astrophysical data as if the errors were statistical and gaussian.

For this reason, we employ an additional procedure, which we call
``concordance,'' to evaluate the observational constraints.   We identify
models as passing the observational constraints if they lie within $2\sigma$ of
each individual constraint. (We do not consider the joint probabilities
spanning two or more observations.) We allow a generous range for systematic
errors.  Not only does this procedure provide a reliable picture of current
constraints, but there is the added advantage that it renders transparent which
observations are most important in delimiting the range of currently allowed
parameters and which future  observations will be most influential. 

One might characterize the difference between the two approaches as follows:
two identical and observationally independent constraints on a single parameter
do not change, at all, the allowed range of that parameter in the  concordance
analysis, but do reduce the range in a maximum  likelihood approach. Clearly,
the concordance approach is too conservative if the errors are known to be
gaussian.  However, in the current case, systematics dominate and the
correlations between errors in different observations is unclear. In these
circumstances, as discussed in the Appendix,  the maximum likelihood approach
can produce  seriously misleading results. One should be especially watchful
and examine closely situations where the concordance and maximum likelihood
analyses strongly disagree.  Maximum likelihood has the undesireable feature
that it will seek a compromise among conflicting data.   Hence, we advocate the
more conservative concordance approach, supplemented with comparison to   a
full maximum likelihood estimator for the  constraints assuming gaussian
errors.

We classify the observations by red shift.  At low red shift, $z \ll 1$, the
constraints are due to:  the Hubble constant; age of the universe; baryon
density; x-ray cluster abundance; shape of the mass power spectrum; and bulk
flow. At intermediate red shift, $z \sim 1$, the constraints are due to:  type
1a supernovae; evolution of the x-ray cluster abundance;  gravitational
lensing; Lyman-$\alpha$ forest.  At high red shift, $z \gg 1$, the constraints
are due to:  the fluctuation amplitude and spectral tilt based on the large
angle CMB temperature anisotropy (COBE); small angle CMB temperature
anisotropy.  The high red shift CMB constraint due to COBE, along with the set
of low red shift results, serve as the strongest, most reliable constraints. 
These constraints will be shown to dominate the boundary of the allowed
parameter range.  The intermediate red shift constraints, due to SNe and the
evolution of x-ray clusters, are rapidly reaching the point where they impact
the range of cosmological models.  The small angle CMB measurements, as well,
are soon to yield prime cosmological information.  We will consider each of
these constraints in turn.

The observational constraints used to restrict the quintessence parameter space
are listed in the following subsection.  The low red shift, and COBE-based high
red shift constraints compose the core set of concordance tests of our
cosmological models.  The remaining intermediate and high red shift constraints
are less certain at the present, although they offer the promise of powerful
discrimination between models in the near future.

\subsection{Low Redshift}
\label{lowzsection}

{\bf H}:  The Hubble constant has been measured through numerous techniques
over the years.  Although there has been a marked increase in the precision of
extragalactic distance measurements, the accurate determination of $H$ has been
slow.  The ${\rm H_0}$ Key Project (\cite{Freedman98}), which aims to measure
the Hubble constant to an accuracy of $10 \%$, currently finds $H = 73 \pm 6
({\rm stat}) \pm 8 ({\rm sys})$ km/s/Mpc; the method of type 1a supernovae
gives  $H = 63.1 \pm 3.4 ({\rm internal}) \pm 2.9 ({\rm external})$ km/s/Mpc
(\cite{HamuyEtAl}); a recent measurement based on the Sunyaev-Zeldovich effect
in four nearby clusters gives $H=54 \pm 14$ km/s/Mpc (\cite{MyersSZE97});
typical values obtained from gravitational lens systems are $H \sim 50-70$
km/s/Mpc with up to $\sim 30\%$ errors
(\cite{Falco97,Keeton97,Kundic97,Schechter97}).  Using surface brightness
fluctuations to calibrate the  bright cluster galaxy Hubble diagram, a
far-field value $H = 89 \pm 10$ km/s/Mpc, out to $\sim 11,000$ km/s, has been
obtained (\cite{LauerEtAl}). Clearly, convergence has not yet been reached,
although  some methods are more prone to systematic uncertainties.  Based on
these diverse measures, our conservative estimate for the Hubble parameter is
$H = 65 \pm 15$~km/s/Mpc at effectively the $2 \sigma$ level. While knowledge
of $H$ would certainly be a decisive constraint, our current uncertainty will
not prevent us from narrowing the field of viable cosmological models.

${\bf t}_{\bf 0}$:  Recent progress in the dating of globular clusters and the
calibration of the cosmic distance ladder has relaxed the lower bound on the
age of the universe.  We adopt  $t_0 \ge 9.5$ Gyr as a $95 \%$  lower limit
(\cite{Chaboyer98,Salaris98}), although we note that some arguments
(\cite{Paczynski99}) suggest a 10\% higher limit is more appropriate.

{\bf BBN}:  Recent observations of the deuterium abundance by Burles and Tytler
yield $D/H = 3.4 \pm 0.3 ({\rm stat}) \times 10^{-5}$
(\cite{BurlesTytler97a,BurlesTytler97b,BurlesTytler97c}).  If this value
reflects the primordial abundance, then big bang nucleosynthesis (BBN; for a
review see \cite{SchrammTurner98} and references therein) with three light
neutrinos gives $\Omega_b h^2 = 0.019 \pm 0.002$, where the $1\sigma$ error
bars allow for possible systematic uncertainty. While other observations
suggest that the abundance varies on galactic length scales where it is
expected to be uniform, suggesting that heretofore unknown processes may be
processing the deuterium (\cite{JenkinsEtAl}), we will adopt the hypothesis
that the  cosmological abundance has been ascertained by the measurements of
Burles \& Tytler.

{\bf BF}:  Observations of the gas in clusters have been used to estimate the
baryon fraction (compared to the total mass) to be $f_{gas}=(0.06 \pm
0.003)h^{-3/2}$ (\cite{Evrard97}; also see \cite{WNES93,FHP97}).  The stellar
fraction is estimated to be less than $20\%$ of the gas fraction, so that
$f_{stellar} = 0.2 h^{3/2} f_{gas}$.  Next, simulations suggest that the baryon
fraction in clusters is less than the cosmological value by about $10\%$
(\cite{LubinEtAl96}) 
representing a depletion in the abundance of baryons in
clusters by a factor of $0.9 \pm 0.1$. Hence, the cosmological baryon fraction
($\Omega_b/\Omega_m$) is estimated to be $f_{baryon} = (0.067 \pm 0.008)
h^{-3/2} + 0.013$ at the $1\sigma$ level.  Using the observed baryon density
from BBN, we obtain a constraint on $\Omega_m$:
\begin{equation}
\Omega_m = \frac{0.019 h^{-2}}{0.067 h^{-3/2} + 0.013}\Big(1 \pm 0.32\Big)
\end{equation}
at the $2\sigma$ level. For $h=0.65$,  this corresponds to a value of
$\Omega_m=0.32 \pm 0.1$.

The baryon fraction has also been estimated on smaller scales such as galaxy
systems (see \cite{McGaugh} and references therein). However, the relationship
between the local baryon fraction on those scales and the global baryon
fraction is uncertain and currently beyond the power of numerical study.

${\bf \sigma}_{\bf 8}$:  The abundance of x-ray clusters at $z=0$ provides a
model dependent normalization of the mass power spectrum at the canonical $8
h^{-1}$~Mpc scale.  The interpretation of x-ray cluster data for the case of
quintessence models has been carried out in detail in \cite{Wang98}, in which
case the constraint is expressed as
\begin{eqnarray} \label{clusconstraint}
\sigma_8 \Omega_m^\gamma &=& (0.5 - 0.1 \Theta) \pm 0.1 
\end{eqnarray}
where the error bars are $2\sigma$, with
\begin{eqnarray}
\gamma &=& 0.21 - 0.22 w + 0.33 \Omega_m + 0.25 \Theta \cr
\Theta &=& (n_s - 1) + (h - 0.65).
\end{eqnarray}
This fitting formula is valid for the range of parameters considered in this
paper.

Perhaps the two most important constraints on the mass power spectrum  at this
time are the COBE limit on large scale power and the  cluster abundance
constraint which fixes the power on  $8 h^{-1}$~Mpc scales.  Together, they fix
the spectral index and  leave little room to adjust the power spectrum to
satisfy other tests.

\placefigure{fig:shape}

{\bf Shape}:  If light traces mass with a constant bias factor on large scales,
then the deprojected APM galaxy cluster data (\cite{Peac97comm}) can be used to
constrain the shape of the underlying mass power spectrum.  The bias factor is
defined as $b^2 \equiv P_{APM}/P_{mass}$, the ratio of the APM to mass power
spectra on a given scale; we assume the bias represents a constant, quadratic
amplification of the clustering power of rare objects over the density field on
large scales (\cite{Kaiser86}).  Hence, in keeping with our spirit of
conservativism, we restrict our attention to wavenumbers which are well within
the linear regime, using only the seven lowest frequency bins (as given in
\cite{Peac97}) for scales above $8$ Mpc/h.  The shape constraint consists of
the requirement that the mass power spectrum fit the seven APM data points with
$b \ge 1$ and a reduced $\chi^2 \le 2.0$, corresponding to a confidence level
of 95\%.  In effect, our shape test depends also on the power spectrum
amplitude. The lower bound on the bias is due to the assumption that the
bright, luminous APM objects are preferentially formed in highly overdense
regions (\cite{Davis85,Bardeen86,Kaiser86,OstrikerCen98,OstrikerBlanton98}). 
(Although the bias may be very large at the time of formation, simple arguments
indicate that by the present time, $b$ may have decreased to no lower than
unity. See \cite{Fry86,TegmarkPeebles98}). The consequences of $b < 1$ will be
discussed.  While we give no upper bound on $b$, almost all the best fitting
values for concordance models fall within the rough (model-dependent) upper
bound estimate of $b \lesssim 1.5$ based on higher-order statistics of the APM
data set and current theoretical {\it ab initio} modeling  (\cite{GaztanagaFrieman94,OstrikerCen98,OstrikerBlanton98}).   Our
computations show that the popular ``shape parameter'' $\Gamma \equiv \Omega_m
h$ is not an accurate description of the goodness of fit to the APM data, given
the variety of models that we consider here, since the amplitude of the power
spectrum is characterized by other combinations of parameters, including $w$.
As illustrated in Figure \ref{fig:shape}, five sample models with $\Gamma$
ranging from 0.20 to 0.52 all pass our shape test based on a $\chi^2$ analysis.

{\bf Velocity Field}:   The large-scale velocity field has long been used as a means to
probe the background and fluctuation matter density field. One method is to compare
peculiar velocity data obtained from distance indicators, such as Tully-Fisher, and
from red shift surveys, in order to estimate $\Omega_m$, modulo an assumption about
biasing. Through this method, the quantity $\beta \equiv f(\Omega_m)/b$ is obtained,
where  $f\equiv d\ln\delta_m/d\ln a \sim \Omega_m^{0.6}$ and $b$ is a linear bias
parameter. A variety of recent results (\cite{DNW,WEA,WS,DCEA}) find $\beta \sim 0.5 -
0.6$. For a bias not too different from unity, these results suggest a low matter
density. In the lack of a more complete understanding of bias, however, this method
cannot be transformed into a rigorous constraint on $\Omega_m$. As well, it remains to
be understood why similar approaches which compare the density fields obtained from
distance indicators and red shift survey data, along the lines of the POTENT method
(\cite{POTENTalgo}),  generally obtain higher values, {\it e.g.} $\beta \sim 0.9$
(\cite{Sigad98}). Another method is to compare observations with the predicted velocity
field within the context of a particular cosmological model. Results based on the Mark
III (\cite{Zaroubi97}) and SFI (\cite{Zehavi98,Freudling99}) catalogs yield the
constraint $f^2 P(k) = (4.8 \pm 1.5) \times 10^3\, ({\rm Mpc/h})^3$ and $(4.4 \pm 1.7)
\times 10^3\, ({\rm Mpc/h})^3$ at $k=0.1\,{\rm h/Mpc}$, respectively. Extrapolating to
smaller scales, within the class of $\Lambda$CDM models, the SFI constraint can be
recast as $\sigma_8 f = 0.82 \pm 0.12$. Due to the discrepancy with the cluster
abundance constraint, we are hesitant to apply this recent result until further
analysis reinforces its conclusions.

\placefigure{fig:bulkv}

The bulk flow on the largest scales provides another method to do cosmology
with the velocity field. In Figure~\ref{fig:bulkv} we compare the predictions
of a set of QCDM models with observation. The large sample variance on the
Maxwellian distributed velocity field means that consistency requires the
observations lie below the upper $95\%$CL bulk velocity. (The lower $95\%$CL
bulk velocity is very small, so we may effectively treat this constraint as an
upper bound.) A measurement near or above the swath of predicted curves could
serve as a strong indicator of the cosmology.  For comparison, we also show the
measured bulk velocities of \cite{Dekel98},  \cite{Giovanelli98}, and
\cite{LP94}.  At present, the diversity of measurements, as displayed in 
Figure~\ref{fig:bulkv}, dilutes the strength of the constraint resulting from
the comparison of the velocity dipole with the CMB dipole.

\subsection{Intermediate Redshift}
\label{midzsection}

{\bf SNe}:  Type 1a supernovae are not standard candles, but empirical
calibration of the light curve - luminosity relationship suggests that the
objects can be used as distance indicators.  There has been much progress in
these observations recently, and there promises to be more.  Hence, a
definitive constraint based on these results would be premature.  However, we
examine the recent results of the High-Z Supernova Search Team (HZS:
\cite{Riess98,Garnavich98}) and the Supernova Cosmology Project (SCP:
\cite{Perlmutter98}) to constrain the luminosity distance - red shift
relationship in quintessence cosmological models.  We have adopted the
following data analysis procedure:  we use the supernova data for the shape of
the luminosity - red shift relationship only, allowing the calibration, and
therefore the Hubble constant, to float; we excise all SNe at $z < 0.02$ to
avoid possible systematics due to local voids and overdensities; for SNe at $z>
0.02$, we assume a further uncertainty, added in quadrature, corresponding to a
peculiar velocity of $300 $km/s in order to devalue nearby SNe relative to the
more distant ones (for the SCP data, a velocity of $300 $km/s has already been
included).  There is substantial scatter in the supernovae data, as seen in
Figure \ref{fig:double_SNe}.  The scatter is so wide that no model we have
tested passes a $\chi^2$ test with the full SCP data set; using  a reduced set,
Fit C, argued by the SCP as being more reliable \cite{Perlmutter98},  a finite
range of models do pass the $\chi^2$ test and the  range is comparable to the
range obtained by the $\chi^2$ test using the HZS data set. To gauge the
current situation, we will report both $\chi^2$ tests and maximum likelihood
tests; to be conservative, we use the largest boundary  (the $\chi^2$ test
based on HZS data using MLCS analysis) for our concordance constraint.

\placefigure{fig:double_SNe}

{\bf Cluster Evolution}:  The abundance of rich clusters --- objects presumed
to have formed from high density peaks drawn from the exponential tail of an
initially Gaussian perturbation distribution -- can be used to constrain the
amplitude of the mass power spectrum at intermediate red shift.   The current
observations have been converted into a number density of clusters above a
certain mass threshold $M_{1.5}$, defined to be the mass within the comoving
radius $R_{com}=1.5 h^{-1}$~Mpc. For the models of interest, the abundance
evolves approximately as a power law for $0<z<1$: \begin{equation}
n(>M_{1.5},z) \propto 10^{ A(M_{1.5}) z}   \end{equation} (see \cite{Wang98}). 
The bigger $A(M_{1.5})$ is, the weaker the evolution is, implying low
$\Omega_m$ and $w$.  Since the measurements  (summarized in \cite{Bahcall97})
are still in the preliminary stage, we adopt $A(M_{1.5}=8\times
10^{14}h^{-1}{\rm Mpc})= -1.7^{+1.5}_{-2.1}$ as a conservative, $2\sigma$
limit. Similar tests have been applied in the context of $\Lambda$CDM and open
models (\cite{Bahcall97,Carlberg97}), for which the results are model
dependent.

{\bf Lensing Counts}:  The statistics of multiply imaged quasars, lensed by
intervening galaxies or clusters, can be used to determine the luminosity
distance - red shift relationship, and thereby constrain quintessence
cosmological models.  There exists a long literature of estimates of the
lensing constraint on $\Lambda$ models (e.g. from \cite{TOG84} to
\cite{Falco98}).  In one approach, the cumulative lensing probability for a
sample of quasars is used to estimate the expected number of lenses and
distribution of angular separations.  Using the Hubble Space Telescope Snapshot
Survey quasar sample (\cite{HSTSSS}), which found four lenses in 502 sources,
Maoz-Rix (\cite{MaozRix93}) arrived at the limit $\Omega_\Lambda \lesssim 0.7$
at the $95\%$ CL.   In a series of studies, similar constraints have been
obtained using optical (\cite{Kochanek95,Kochanek96}) and radio lenses
(\cite{Falco98}).  Waga and collaborators (\cite{TorresWaga96,WagaMiceli98})
have generalized these results, finding that the constraint weakens for larger
values of the background equation-of-state, $w > -1$ (as noted earlier by
\cite{RatraQuillen}).  In our evaluation of the constraint based on the HST-SSS
data set, we find that the $95 \%$ confidence level region is approximately
described by $\Omega_Q \lesssim 0.75 + (1+w)^2$, until the inequality is
saturated at $w = -1/2$, consistent with \cite{TorresWaga96,WagaMiceli98}.   In
principle, this test is a sensitive probe of the cosmology; however, it is
susceptible to a number of systematic errors (for a discussion,  see
\cite{MRT97,CK99}).  Uncertainties in the luminosity function for source and
lens, lens evolution, lensing cross section, and dust extinction for optical
lenses, threaten to render the constraints compatible with or even favor a low
density universe over $\Omega_m=1$.  Taking the above into consideration, none
of the present constraints on quintessence due to the statistics of multiply
imaged quasars are prohibitive:  models in concordance with the low-$z$
constraints are compatible with the lensing constraints.

{\bf Ly-${\bf\alpha}$} (and other mass power spectrum measurements at moderate
red shift, $1<z<10$):  The Ly-$\alpha$ forest has been used as the basis of a
number of cosmological probes.  Most recently, the effect of the local mass
density in the intergalactic medium on the Ly-$\alpha$ optical depth
(\cite{Hui98,Croft98a}) has been used to estimate the mass power spectrum at a
red shift of $z=2.5$ (\cite{Croft98b,Weinberg98}).  This is a good pedagogical
example to study what can and cannot be learned from studies at moderate red
shift $1<z<10$.   In Figure \ref{fig:lya} we show the linear mass power
spectrum today (upper panel) and at red shift $z=2.5$ (lower panel) for the
representative best-fit models discussed in Table I. The mass power spectra in
the upper panel all satisfy COBE normalization at large scales and the cluster
abundance constraint on $\sigma_8$ on 8~$h^{-1}$~Mpc scales. Since the models
have already passed the constraints on 8~$h^{-1}$~Mpc scales and higher, one
might hope that tests of the linear power spectrum at smaller scales might
further distinguish the models. However, smaller scales correspond to the
non-linear growth regime  where effects like scale-dependent bias make it
difficult to compare observations to the linear power spectrum.  Measurements
of  the Lyman-$\alpha$ forest are promising because they probe the  power
spectrum on smaller scales at a red shift before non-linearities develop and,
hence, enable direct comparison to the linear power spectrum.   We draw the
reader's attention to the fact that,  in converting to  $z=2.5$, we have made a
model-dependent rescaling of the abscissa  so that the units are those of
velocity, which then allows direct comparison to the data. In the upper ($z=0$)
panel, the models differ substantially on large scale but appear to converge on
small scales.  Projecting back to $z=2.5$ and rescaling,   one might hope that
the models are distinct due to the differing growth  functions. However,
instead, examples with the same spectral  tilt and $\Omega_m H_0/H(z)$ (as is
demonstrated inadvertently by our representative models)  nearly overlap
everywhere, making  discrimination very difficult.  When $\Omega_m H_0/H(z)$ is
fixed, as shown in the figure,  then, we note,  the Lyman-$\alpha$ measurements
can be used to determine the tilt, $n_s$.

\placefigure{fig:lya}

The current data appears to favor $n_s=1$. There is some hope that improved 
limits can discriminate among models with different  tilt, but  determining
other parameters, and especially discriminating between cosmological constant
and quintessence on this basis, appears hopeless because the predictions of
various models converge, as shown in the lower  panel. Of course, this
conclusion applies not only to Lyman-$\alpha$ forest measurements, but any
approach that measures the mass power spectrum at moderate red shift.

Further applications of the Ly-$\alpha$ forest, such as the abundance of damped
Ly-$\alpha$ absorbers (\cite{Gardner97}) and correlations in the lines of sight
at red shifts $z \sim 2 - 4$ (\cite{McDonald98,HuiEtAl98}) have been developed
as tests of geometry and expansion history,  although no substantial
constraints have as yet been obtained.  

There are a number of observational probes which are sensitive to the
cosmology, but which have not yet matured into critical tests.  We list some of
these tests which may prove to be powerful constraints in the near future.
Measures of the abundance of objects, similar to the cluster evolution
constraint, can be used to gauge the growth of structure. Observations of
galaxies formed as early as $z \gtrsim 3$ (\cite{Steidel98}) have been
interpreted, on the basis of a Press-Schechter formalism
(\cite{PressSchechter74}), to suggest that among a family of CDM cosmologies,
flat, low-density models best satisfy the constraint (\cite{MoFukugita96}). 

Finally, it has been  proposed to use the statistics of gravitational lens arcs
produced by intermediate red shift clusters as a means to distinguish
cosmological models (\cite{WuMao96,Bartelmann98}).  These are potentially
extremely powerful tests,  insofar as the rare, high density fluctuations
reflect the underlying cosmology. In particular, the lens arcs statistics are
exponentially  sensitive to the growth function, which differs for $\Lambda$
versus quintessence models and, hence,  has the potential  of distinguishing
the two scenarios.  However, results based on  numerical simulations are not
yet capable of resolving the core of clusters accurately enough to provide
reliable limits (\cite{Wambsganss98}) from the theoretical side.
 
\subsection{High Redshift}
\label{highzsection}

One of the most powerful cosmological probes is the CMB anisotropy, an imprint
of the recombination epoch on the celestial sphere.  The large angle
temperature anisotropy pattern recorded by COBE can be used to place two
constraints on cosmological models.

{\bf COBE norm}:  The observed amplitude of the CMB power spectrum is used to
constrain the amplitude of the underlying density perturbations.   We adopt the
method of \cite{BunnWhite97} to normalize  the power spectrum to COBE. As we
use a modified version of CMBFAST (\cite{CMBFAST}) to compute the CMB
anisotropy spectra, this normalization is carried out automatically.) We have
verified that this method, originally developed for $\Lambda$ and open CDM
models, can be applied to the quintessence cosmological models considered in
this work (\cite{RahulDave}). Of course, there is uncertainty associated with
the COBE ``normalization'':  the $2\sigma$ uncertainty in rms quantities is
approximately $20\%$ (see footnote \#4 in \cite{BunnWhite97}), which
conservatively allows for statistical errors, as well as the systematic
uncertainty associated with the differences in the galactic and ecliptic frame
COBE map pixelizations, and potential contamination by high-latitude
foregrounds({\it e.g.} \cite{GorskiNorm}).  

${\bf n}_{\bf s}$:  COBE has been found to be consistent with a $n_s=1.2 \pm
0.3$ spectral index (\cite{Gorski96,Hinshaw96}), but this assumes the only
large angular scale anisotropy is generated via the Sachs-Wolfe effect on the
last scattering surface.  This neglects the baryon-photon acoustic
oscillations, which produce a rise in the spectrum, slightly tilting the
spectrum observed by COBE.  In general, the spectral index determined by
fitting the large angular scale CMB anisotropy of a quintessence model, which
is also modified by a late-time integrated effect, to the shape of the spectrum
tends to overestimate the spectral tilt.  For example, analysis of a class of
CDM models (\cite{Hancock98})  ($\Lambda$CDM and SCDM, a subset of the models
considered here) finds a spectral tilt $n_s = 1.1 \pm 0.1$.  We conservatively
restrict the spectral index of the primordial adiabatic density perturbation
spectrum, with $P(k) \propto k^{n_s}$, to lie in the interval $n_s \in [0.8,
1.2]$.  Note that inflation generically predicts $n_s \sim 1$, with $n_s$
slightly less than unity preferred by inflaton potentials which naturally exit
inflation.

{\bf Small Angle CMB}:  Dramatic advances in cosmology are expected in the near
future, when the MAP and Planck satellites return high resolution maps of the
CMB temperature and polarization anisotropy.  When the measurements are
analyzed, we can expect that the best determined cosmological quantities will
be the high multipole $C_\ell$ moments, such that any proposed theory must
first explain the observed anisotropy spectrum.  At present, however, there is
ample CMB data which can be used to constrain cosmological models.

\placefigure{fig:cmb}

We take a conservative approach in applying the small angular scale CMB data as
a model constraint.  Our intention is to simply determine which quintessence
models are consistent with the ensemble of CMB experiments, rather than to
determine the most likely or best fitting model; examining
Figure~\ref{fig:cmb}, the error bars are clearly so large that a `best fit' has
little significance.   We have restricted our attention to that subset of 
published CMB experimental data which satisfies the following objective
criteria: multi-frequency; positive cross-correlation with another experiment;
careful treatment of foregrounds. (While we advocate these criteria, our
conclusions are not strongly sensitive to this selection of data.) Hence, we
use the bandpower estimates from COBE (\cite{COBE96}), Python (\cite{PYTHON}),
MSAM (\cite{MSAM}), QMAP (\cite{QMAPa,QMAPb,QMAPc}), Saskatoon (\cite{SK}), CAT
(\cite{CAT}), and RING5M (\cite{RING}) experiments as the basis of the small
angle CMB cosmological constraint.   Figure \ref{fig:cmb} shows the bandpower
averages at the effective multipole number, $l_e$, with several QCDM models for
comparison.  We apply a simple $\chi^2$ test with the predicted bandpower
averages, $\delta T_{l_e}$ (defined in \cite{Bond95}).  (A compilation of
bandpower averages and window functions is available from 
Knox\footnote{www.cita.utoronto.ca/$\sim$knox/radical/bpdata.html}
and 
Tegmark\footnote{www.sns.ias.edu/$\sim$max/cmb/experiments.html}.)  Since the reported bandpower errors are typically
not Gaussian distributed, treating the $\delta T_{l_e}$ as a Gaussian random
variable can introduce a bias in the estimation of the quality of agreement. 
In this case, we also consider a $\chi^2$ test in the quantity  $\ln(\delta
T_{l_e}^2)$, following \cite{Bond98}.

Figure~\ref{fig:cmb}  illustrates features relevant to limits on the spectral
index, $n_s$.  The COBE limit on the spectral index is reported to be $n_s=1.2
\pm 0.3$ (\cite{Gorski96,Hinshaw96})  based on comparison between data and 
standard CDM models. In the Figure, though, it is apparent that there is
negligible difference in the low angular scale predictions despite a range of
$\Delta n =0.2$.  The difference  in the large angular scale integrated
Sachs-Wolfe contribution  compensates for the difference in spectral index.
Hence, as will be addressed in a later paper, the  COBE limit on $n_s$ is
somewhat expanded when QCDM is included.

\section{Concordance Results}
\label{resultsection}

We have evaluated the cosmological constraints for the set of quintessence
models occupying the five dimensional parameter space: $w, \,\Omega_m, \,
\Omega_b, \,h, \,n_s$.  The results are best represented by projecting the
viable models onto the $\Omega_m-h$ and $\Omega_m-w$ planes.

The concordance region due to the suite of low red shift constraints, including
the COBE normalization and tilt $n_s$, are displayed in Figures \ref{fig:om-h},
\ref{fig:om-w}.  Each point in the shaded region represents at least one model
in the remaining three dimensional parameter space which satisfies the
observational constraints.  

\placefigure{fig:om-h}

In Figure \ref{fig:om-h}, the boundaries in the $\Omega_m$ direction are
determined by the combined BBN and BF constraints as a function of $h$, while
$h$ is only restricted by our conservative allowed range and the age
constraint.  The age does not impact the $\Omega_m - h$ concordance region,
since for the allowed values of $\Omega_m$ and $h$, there is always a model
with a sufficiently negative value of $w$ to satisfy the age constraint.
Relaxing either the BBN or BF constraint would raise the upper limit on the
matter density parameter to allow larger values of $\Omega_m$.  This requires a
simultaneous reduction in the spectral index, $n_s$, in order to satisfy both
the COBE normalization and cluster abundance. 

\placefigure{fig:om-w}

In Figure \ref{fig:om-w}, the upper and lower bounds on $\Omega_m$ are again
determined by the combination of BBN, BF, and $h$.  The lower bound on
$\Omega_m$ due to the combination of the BBN and BF constraints can be relaxed
if we allow a more conservative range for the baryon density, such as $0.006 <
\Omega_b h^2 < 0.022$ (\cite{Lev98,OSW99}). However, the constraints due to
$\sigma_8$ and the shape of the mass power spectrum take up the slack, and the
lower boundary of the concordance region is relatively unaffected. The lower
bound on $\Omega_m$ near $w=-1$ is determined in part by the shape test; the
mass power spectrum in a model with low $\Omega_m$ and  strongly negative $w$
is a poor fit to the shape of the APM data, based on a $\chi^2$-test.   This
constraint on   models near $w=-1$ is relaxed if we allow  anti-bias ($b<1$),
although $b<1$ is strongly disfavored on a theoretical basis. At the other end,
for $w \gtrsim -0.6$, the lower bound on $\Omega_m$ is determined by the
combination of the upper bound on the spectral index, and the x-ray cluster
abundance constraint on $\sigma_8$. If we further restrict the bias to
$b<1.5$,  a small group of models at the upper right corner with $w \gtrsim
-0.2$ and $\Omega_m \gtrsim 0.4$ will fail the shape test.  

We see that models occupying the fraction of the parameter space in the range
$-1 \le w \lesssim -0.2$ and $0.2 \lesssim \Omega_m \lesssim 0.5$ are in
concordance with the basic suite of observations, suggesting a low density
universe.   It is important to note that the set of viable models spans a wide
range in $w$; the concordance region is not clustered around $w=-1$, or
$\Lambda$, but allow such diverse behavior as $w \sim -1/3$.   However, the
case $w=0$, which can result from the scaling exponential potential
(\cite{RatraPeebles88,PeeblesRatra88}), is clearly in contradiction with
observation:  the $\Omega_m$ required by the x-ray cluster abundance constraint
is incompatible with the matter density parameter allowed by the BF and BBN
constraints.  Hence, the models with $w=0$ explored in
\cite{FerreiraJoyce97,FerreiraJoyce98} are not viable.

\placefigure{fig:om-w.mle}

We have taken the attitude in our work that current observational uncertainties
are dominated by systematic errors, so that a conservative method of combining
observational constraints is by concordance.  We  apply the  $2\sigma$ limits
for each individual observation to pare down the viable parameter range.
However, it is interesting to compare this to what  a naive maximum likelihood
estimate  (treating the errors as gaussian) would give. In
Figure~\ref{fig:om-w.mle} we show the $2\sigma$ contour in the $\Omega_m - w$
plane, where the remaining parameters have been marginalized.  This parameter
region is only slightly smaller than  that resulting from concordance.  It is
reassuring that this technique yields approximately the same result,  although
one should be cognizant of some of the pitfalls of both  methods as discussed
in the Appendix. In using the maximum likelihood technique, we lose some of our
ability to identify the constraint dominating a particular portion of the
contour.  However, we show that by lifting the shape test, the constraint
relaxes on the range of $\Omega_m$ allowed for models with $w$  closer to $-1$,
including $\Lambda$CDM.  For the sake of argument, the `best fit' models, where
the likelihood is maximized, are also shown. We see more clearly that the shape
test drives the preferred models away from $w = -1$, towards $w \sim -1/2$. 

\placefigure{fig:om-w.shrink}

The concordance approach offers no such `best fit', as it contains no procedure
for weighting or combining data.  However, in Figure~\ref{fig:om-w.shrink} we
carry out the exercise of artificially shrinking all the error bars, to find
the last remaining models.  This is equivalent to imagining that all
measurements have accurately determined the intended quantity, but overstated
their uncertainties.  This procedure narrows down to the same set of models,
$\Omega_m \sim 0.33$, but is not driven as strongly by the shape test as is the
maximum likelihood procedure.

\placefigure{fig:om-wsnemle}

The most potent of the intermediate red shift constraints is due to type 1a
supernovae, which we present in Figure~\ref{fig:om-wsnemle}. In addition to the
SCP results, the HZS group has presented two different analyses of their
catalog of SNe, based on multi-color light curve shapes (MLCS) and template
fitting; hence we show three SNe results. Carrying out a maximum likelihood
analysis, all three give approximately the same result for the location of  the
$2\sigma$  bound, favoring concordant models with low $\Omega_m$, and very
negative $w$.  It is interesting to observe that the SNe bound is consistent
with the core of the low red shift concordance region, displayed earlier in
Figure~\ref{fig:om-w.shrink}.  Based on the SCP maximum likelihood analysis,
\cite{Perlmutter99} have reported a limit $w\le -0.6$ at the 1$\sigma$ level. A
$\chi^2$ analysis of the same data gives a somewhat different result:  the Fit
C SCP data and the HZS data sets give comparable, although weaker, results to 
the likelihood analysis.  In the spirit of conservativism, we have used the
weakest bound which we can reasonably justify. Hence, for the concordance
analysis, we use the $2\sigma$ contour resulting from a $\chi^2$ test. 

\placefigure{fig:lensing}

The statistical rate of gravitational lensing provides a counter to the trend
towards low matter density.  In Figure~\ref{fig:lensing} we present the results
of our analysis of the HST-SSS lenses. Other groups have come to similar
conclusions, based on this and other lens surveys.   Our results are in
excellent agreement with \cite{TorresWaga96,WagaMiceli98}, as well as the more
sophisticated analyses carried out by
\cite{MaozRix93,Kochanek95,Kochanek96,Falco98} for $\Lambda$CDM. In general,
there are fewer lenses observed than expected based on the volume-red shift
relation for a low density, $\Lambda$-dominated universe.  The disparity
between theory and observation is reduced as the matter density increases, or
as $w$ increases.

\placefigure{fig:om-w.cluster}

We have evaluated the x-ray cluster evolution constraint using the observed
abundance of rich clusters at $z\sim 1$.  This test constrains the amplitude of
mass fluctuations and the rate of perturbation growth.  While it has been
argued that this reduces to a bound on $\sigma_8$ for $\Lambda$CDM models
(\cite{BahcallFan98}), it has been shown that the bound depends on $w$ for QCDM
scenarios by \cite{Wang98}. In Figure \ref{fig:om-w.cluster} we show the
consequence of the cluster evolution constraint on the concordance models in
the $\Omega_m-w$ plane; at this  stage, the early formation of structure
implied by the observations argues against concordant quintessence models with
an equation-of-state $w \gtrsim -0.3$.  When the measurements comprising this
constraint improve, we can expect a much more stringent result.  Considering
the hypothetical situation that future observations successfully reduce the
systematic uncertainty to the present $1\sigma$ level, the constraint boundary
would shift to  the small region with $-0.8 \lesssim w \lesssim -0.5$ and $0.25
\lesssim \Omega_m \lesssim 0.3$.

We have evaluated the high red shift constraint due to the select ensemble of
CMB anisotropy measurements, using the COBE, Python (\cite{PYTHON}), MSAM
(\cite{MSAM}), QMAP (\cite{QMAPa,QMAPb,QMAPc}), Saskatoon (\cite{SK}), CAT
(\cite{CAT}), and RING5M (\cite{RING}) results.  Based on a $\chi^2$ test in
$\delta T_l$, the set of concordant models projected down to the $\Omega_m-h$
and $\Omega_m-w$ planes is unchanged from the low red shift concordance region
at even the $1\sigma$ level.  This ``null'' result from the CMB should not be
too surprising; the current observational data is capable only of discerning  a
rise and fall in power  in the $C_\ell$ spectrum across $\ell\sim 100-300$. 
The results are unchanged if we include additional current  CMB results, or use
a $\chi^2$ test in $\ln(\delta T_l^2)$, as suggested in \cite{Bond98}. Rather,
we must wait for near-future experiments which have greater $\ell-$coverage,
{\it e.g.} BOOMERANG, MAT, and MAXIMA, which are expected to significantly
reduce the uncertainties. 

Since the submission of this manuscript, the data from the MAT (Miller et al,
1999)   and BOOMERANG (Mauskopf et al, 1999) experiments have been released.
However, neither significantly changes  our results.

\placefigure{fig:ult_concord}

Thus far we have applied the low red shift constraints in sequence with one of
the other intermediate or high red shift constraints.  It is straight forward
to see how the combined set of constraints restrict the quintessence parameter
space.  Taking the low red shift constraint region, which is shaped primarily
by the BF, BBN, H, and $\sigma_8$ constraints, the dominant bounds on the
$\Omega_m - w$ plane are then due to SNe and lensing.  The SNe drives the
concordance region towards small $\Omega_m$ and negative $w$; the lensing
restricts low values of $\Omega_m$. Putting these all together, an ultimate
concordance test is presented in Figure~\ref{fig:ult_concord}. We see that the
resulting concordance region in the $\Omega_m - w$ plane is very similar to the
core region obtained in Figure~\ref{fig:om-w.shrink}.  If the present
observations are reliable, we may conclude that these models are the most
viable among the class of cosmological scenarios considered herein.

It is beyond the scope of the present work to determine how well future
observations will determine the values of cosmological parameters in a QCDM
scenario.  However, we are in a position to highlight those observations which
appear well suited to testing the quintessence hypothesis.  Clearly, the first
goal must be to distinguish Q from $\Lambda$ (\cite{Huey98}).  Observations
which measure the growth of structure at intermediate red shifts ($z\sim
0.5-1.0$) are best suited for the purpose.  In this red shift regime, structure
growth is still occurring for the $\Lambda$ model, but has shut off
significantly for the  Q model. The Ly-$\alpha$ determination of the mass power
spectrum amplitude is at too high of a red shift to serve this purpose:  at
$z\sim 2.5$, evolution is still matter dominated for $\Omega_m=0.3$ models with
$w \lesssim -1/3$.  Cluster abundances at $z \sim 0.5 - 1$ and the supernovae
magnitude-red shift relation are better suited to this goal.

The shape of the mass power spectrum may prove to be a strong test of the QCDM
scenario, if the observation of a  turnover in the power spectrum near $k\sim
0.02 - 0.06$~h/Mpc  and a break in the slope at higher wave numbers
(\cite{GaztanagaBaugh})  bears out.  It may prove difficult for the simplest,
scalar field quintessence, or $\Lambda$ for that matter, to generate such a
feature. This is the subject of another investigation (\cite{ZlatevFeature99}).

A test of the tracker quintessence scenario can be made by determining the
change in the equation-of-state.  If the equation-of-state can be measured at
the present and at an earlier epoch, say $z\sim 1$, we can obtain a crude
measure of the slope, $dw/dt$. Trackers have the special property that the
equation-of-state becomes more negative at late times:  $w \to -1$ as $\Omega_Q
\to 1$.  A measurement of $dw/dt > 0$ would argue against tracker quintessence.

More exotic observational tests can be used to discover the presence of a 
quintessence field. For example, if the quintessence field is coupled to the
pseudoscalar $F_{\mu\nu}\tilde{F}^{\mu\nu}$ of electromagnetism as suggested by
some effective field theory considerations (\cite{Carroll98}), the polarization
vector of a propagating photon will rotate by an angle $\Delta \alpha$ that is
proportional to the change of the field value $\Delta Q$ along the path.   CMB
polarization maps can potentially measure the $\Delta \alpha$ from red shift
$\sim 1100$ to now (\cite{LueEtAl98}) and distant radio galaxies and quasars
can provide information of $\Delta \alpha$ from red shift a few to now
(\cite{Carroll98}). If these two observations generate non-zero results, they
can provide unique tests for quintessence and the tracker hypothesis, because
tracker fields start rolling early (say, before matter-radiation equality)
whereas most non-tracking quintessence fields start rolling just recently (at
red shift of a few).

\section{Conclusions}
\label{discussion}

We have applied a battery of tests and constraints to the family of
quintessence cosmological models, determining the range of parameters which are
concordant with observations.  The most reliable constraints are those
resulting from low red shift observations, and the COBE normalization of the
mass power spectrum.  These restrict QCDM models to a narrow range of
parameters, characterized by low matter density, $0.2 \le \Omega_m \le 0.5$,
and negative equation-of-state, $-1 \le w \lesssim -0.2$.  While the
intermediate red shift results are still developing, the implications are very
exciting.   The  SNe observations narrow the range of matter density near
$\Omega_m \sim 0.3 - 0.4$, and force the equation-of-state to $w \lesssim
-0.4$.  While this appears consistent with the core of the low red shift
concordance, the potential for conflict is present if the matter power spectrum
shape test demands $w \gtrsim -1$.  Our results based on low red shift
observations are given by Figures~\ref{fig:om-h}, \ref{fig:om-w}; adding the
supernovae constraints, which are more recent and whose systematic errors have
not been fully tested,  produces the narrower range shown in
Figure~\ref{fig:ult_concord}.

\placefigure{fig:picasso}

To what degree do current uncertainties in the Hubble parameter, the spectral 
tilt and other cosmic parameters obstruct the resolution in  $w$?  To judge
this issue, we have performed an exercise in  which we fix $h=0.65$, $\Omega_b
h^2=0.019$, and we choose the spectral tilt to insure that the central values
of the COBE normalization and the cluster  abundance constraint are precisely
satisfied.  In Figures~\ref{fig:picasso} and \ref{fig:picasso2},   we show how
different constraints restrict the parameter planes.  Note first the long,
white concordance region that remains in the $\Omega_m-w$ plane, which  is only
modestly shrunken compared to the concordance region obtained when current
observational errors are included.  The region encompasses both $\Lambda$ and a
substantial range of quintessence. Hence, current uncertainties in other
parameters are not critical to the uncertainty in $w$.  The figure further
shows how each individual constraint acts to rule out regions of the plane. The
color or  numbers in each patch represent the number of constraints violated by
models in that patch. It is clear that regions far from the concordance region
are ruled out by many constraints. Both figures also show that  the boundaries
due to the constraints tend to run parallel to the  boundary of the concordance
region. Hence, shifts in the values or the uncertainties in these measurements
are unlikely to resolve the uncertainty in $w$ by ruling out one side or the
other --- either the constraints  will remain as they are, in which case the
entire concordance region is  allowed, or the constraints  will shift to rule
out the entire region. 

\placefigure{fig:picasso2}

New measurements not represented in this figure will be needed to 
distinguish $\Lambda$ from quintessence.  At this point, high precision
measurements of the cosmic microwave background anisotropy are the most
promising.

The results presented in this paper apply to quintessence models in which the
equation-of-state is constant or slowly varying with time. In the latter case,
setting $w = \widetilde w$, given in equation (\ref{effect}), gives an excellent
approximation to the observational predictions for a broad class of models.

\placefigure{fig:tracker}

Particularly important classes of quintessence models are tracker  and creeper
fields.  The tracker models are highly appealing  theoretically because they
avoid the ultra-fine tuning of initial conditions required by models with a
cosmological constant or other (non-tracking) quintessence models. An
additional important feature of these  models is that they predict  a definite
relationship between the present day energy density and pressure, which yields
a lower bound on the constant, effective equation-of-state, near $\widetilde
w\sim -0.8$ (\cite{SteinhardtEtAl98}). Note that the effective or averaged
equation-of-state as computed from Eq.~(\ref{effect}) is about 10 per cent
larger than  the value of $w$ today (given in Table I).  In
Figure~\ref{fig:tracker} we add this bound to the low red shift constraints, 
obtaining the concordance region for tracker quintessence. This region retains
the core of our earlier low red shift concordance, and is consistent with the
SNe constraints. Creeper fields occur for the same potentials as trackers, but
the initial energy density exceeds the radiation density at early times. The
consequence is that the field rapidly rolls down the potential, towards a point
which is only mildly (logarithmically) sensitive to the initial conditions,
where it sticks and is effectively frozen with constant potential energy at
very early times.  Hence, the creeper field has an equation-of-state $w=-1$,
and is effectively indistinguishable from a cosmological constant today.

\placefigure{fig:stracker}

In Figure~\ref{fig:stracker} we combine all current observations on tracker
models.  Since these are arguably the best-motivated theoretically, we identify
from  this restricted region a sampling of representative models, listed in
Table~I, with the most attractive region for quintessence models
being $\Omega_m \approx 0.33 \pm 0.05$, effective equation-of-state $w \approx
-0.65 \pm 0.07$ and  $h=0.65 \pm 0.10$ and are consistent with spectral index
$n_s=1$ indicated by the dark shaded region in Figure~\ref{fig:stracker}.
These models represent the best targets for future analysis.   The challenge is
to prove or disprove the efficacy of these models and, if proven,  to
discriminate among them.

\acknowledgements

We wish to thank Neta Bahcall, John Peacock and Michael Strauss for useful
discussions. This research was supported by the US Department of Energy grants
DE-FG02-92ER40699 (Columbia) and DE-FG02-91ER40671 (Princeton).

\appendix
 
Throughout this paper, we have  chosen to judge  models by  combining
observational constraints according to the ``concordance" method in addition to
the  maximum likelihood  estimator (MLE) method.  We have asserted that the
concordance method is  conservative, sometimes giving a more reliable judge of
the situation than the MLE method, especially  when the observational
constraints may be dominated by systematic, nongaussian, and/or correlated
errors.  We advocate using both concordance and MLE methods, as we have done in
this paper, and then analyzing the source of any  discrepancy before
determining which models should be ruled out. Since  it has been commonplace
to  present MLE results alone, we thought it would be useful  to illustrate
some of the pitfalls that can arise. 

For this purpose, we employ a  toy example in which  we have two parameters,
$A$ and $B$, and two independent (observational) constraints,  represented as
the $2\sigma$ regions $C_1$ and $C_2$, which restrict the allowed ranges of
parameters.  This is meant to be a simplification of  our real situation where 
we have a five-dimensional parameter space to analyze quintessence models, and
we have many observational constraints. In our paper, we have tried to
determine constraints in the $\Omega_m$-$w$ plane by projecting, effectively,
from five-dimensions to two.  In our toy model, we imagine projecting onto  the
$A$-axis to determine the constraint on $A$, indicated as a bar along the axis.
Our interest is to compare the concordance region corresponding to $2\sigma$
with the $95\%$CL contour from the MLE method.

\placefigure{fig:case3}

We  present several simple examples in which there is a large disparity between
the concordance and MLE procedures.   In Figures~\ref{fig:case3},
\ref{fig:case1}, we represent a two dimensional parameter space, $A$-$B$,  with
two independent constraint regions, $C_{1,2}$ shown as shaded rectangles.
The projection of the concordance and MLE regions, $C_{conc}$ and $C_{MLE}$
respectively, onto the horizontal axis are shown as thick strips. Although the
following discussion is qualitative, the relative sizes of the projected strips
are correct.

Given the two observations, $C_1$ and $C_2$, the concordance region  for $A$ is
obtained by: (a) finding their intersection in two-dimensions;  and (b)
projecting the two-dimensional intersection $C_{conc}$ onto the $A$-axis to
obtain a bar.   Note that  we do not project first, and then take the
intersection.   This method can lead to gross errors. For example, consider
Figure \ref{fig:case3}, in which the $C_1$ and $C_2$ have no intersection at
all. By our method, the concordance region is properly identified as the null
set, whereas   projecting first and then finding the intersection would
produce  a considerable band of acceptance, a false conclusion.  

According to the MLE method, we are required to know the central value, $\vec
\mu$, of each region, and assume the errors, $\vec \sigma$, are Gaussian.  In
the following, we make the  simplistic assumptions that the likelihood function
for each observation is symmetric about the center of the constraint region,
$C_1$ or $C_2$.  We weight each point $\vec x$ by a Gaussian $f(\vec x,\vec
\mu,\vec \sigma)$ for each of the two constraints, and identify  $C_{MLE}$, the
contour of constant $f_1 f_2$ or $\chi^2$ which contains $95\%$ of the total
probability, $\int_C f_1 f_2 d\vec x = 0.95\int f_1 f_2 d\vec x$. This is a
straightforward procedure.

\placefigure{fig:case1}

{\bf Case 1:} Figure~\ref{fig:case1} illustrates a case where $C_1$ encompasses
$C_2$.  In this situation, the MLE method (indicated by the lower bar) produces
a {\it smaller} acceptance region than the concordance method. If the errors
are truly gaussian and uncorrelated, the conclusion based on MLE is the better
representation of uncertainty.  Note that the concordance region  includes the
MLE region plus an additional range  of $A$; so, the error made with
concordance is to include too much.  However, no model is ruled out by
concordance which  intuition suggests ought not be eliminated.  We call this
``conservative," and our aim is to find a robust, conservative method.

The MLE is not a robust,  conservative method, as can be illustrated by the
same figure.  Suppose that the observations in $C_1$ and $C_2$ are suspected to
be nongaussian or systematic or correlated.   Then, we are clearly mistaken to
rely on the MLE method and eliminate the range of $A$ which lies within the
concordance region but outside the MLE region.  In this example, the difference
is only modest, but if we were combining a number of observations, the MLE
allowed region would be much tinier than the concordance region, and the
error in trusting the MLE method would be more serious. This point is directly
relevant to this paper.

{\bf Case 2:} The observational constraints $C_1$ and $C_2$ intersect in a
small region or, as illustrated in Figure~\ref{fig:case3},  have no
intersection at all. This case is opposite to Case 1 in that the MLE method
produces a {\it larger} acceptance region  than the concordance method.  For
example, the concordance region in Figure~\ref{fig:case3} is the null set,
whereas the MLE  contour suggests a large acceptance region. This is not a case
where the MLE method is being conservative; rather, it is a case where the MLE
method is misleading. Intuition dictates that the observational constraints are
in conflict, and the concordance method reflects this conclusion by producing a
null concordance region. The MLE method, taken at face value, suggests a broad
range of  agreement.  To be fair to the MLE methodology, one is not supposed to
accept the $95\%$ MLE contour at face value. The contour represents a
probability compared to the maximum likelihood point, and one  is supposed to
check that this point is indeed a good fit. In practice, though, this step is
often ignored or discounted. For example, if the maximum likelihood point has a
high value of  $\chi^2$ by the conventional $\chi^2$-test,   this is often
(properly) considered  a problem due to underestimating experimental errors. 
But, as indicated by this example, the  same statistical result can be an
indication that there is a true contradiction between models and data, and that
completely new models need to considered.  Hence, a contradiction between
concordance and MLE methods is a warning to examine closely the cause. 

{\bf Case 3:}   Suppose constraint $C_2$ is obtained by combining many
measurements with small statistical uncertainty but unknown correlated error.
Then, in a MLE analysis, constraint $C_2$ receives undue statistical weight.

If we drop the simplistic assumption that the likelihood function for each
individual constraint is symmetric about the center of the constraint region,
which is rare, other kinds of discrepancies between concordance  and maximum
likelihood can occur.   For example, the MLE region may be  shifted with
respect to the concordance region so that each test allows models which the
other does not.

As a prominent example, the current, highly provocative measurements of type 1a
supernovae by the High-Z Supernova Search Team (HZS:
\cite{Riess98,Garnavich98}) and the Supernova Cosmology Project (SCP:
\cite{Perlmutter98}) exemplify all three cases above. For the HZS data based
on  MLCS (multi-color light curve shape) analysis or for SCP using the Fit C
selected data set \cite{Perlmutter98},  the concordance region is significantly
larger than the MLE acceptance region, as demonstrated by comparing
Figures~\ref{fig:om-wsnemle} and \ref{fig:ult_concord}.  So, relying on the MLE
$95\%$CL region eliminates models which formally pass the absolute $\chi^2$
test at the $95\%$ level.  While statisticians  may argue that the MLE
likelihood estimate is more reliable {\it assuming uniform prior} over the
parameter space, in comparing qualitatively different models (which do not have
uniform prior), the readers should beware that MLE can potentially rule out an
entire  model even if the model agrees at better than the 95\% confidence level
(as judged by $\chi^2$).  Especially at this early stage when better data will
soon be available, we advocate the more cautious, concordance   approach. This
first example is like Case 1.  We have also  noted that  the scatter in the
supernova  red shift - magnitude data  is so wide that no model which  we have
tested passes a $\chi^2$ test with the SCP  full data set.  Hence, this is an
example, like Case 2,  where the   concordance region is null  but the MLE
acceptance region is large. Finally, in a MLE or Fisher matrix analysis which
combines the supernovae measurements with other observations, the fact that 
there are many individual supernovae with small reported uncertainty gives
these measurements heavy statistical weight. However, as the survey teams
admit, the measurement approach is new and there remains the  possibility that
as yet unidentified physical effects cause a systematic, apparent reddening of
the data.  As in  Case 3, results obtained by simple statistical combination of
supernovae data with other  measurements should be viewed cautiously.

Another feature of maximum likelihood analysis, well-known to  practitioners
but perhaps unappreciated by some, is the fact that  the estimation of
parameters in a multi-dimensional fit can be markedly different from the
estimate when marginalizing over some parameters. For example, the range of $w$
spanned by the MLE 95\% confidence region  in the $\Omega_m$-$w$ plane (as
considered by \cite{Perlmutter99} and in this paper) is significantly narrower
than the  MLE 95\% confidence region in the three-dimensional parameter space
including $h$ and significantly broader than the  MLE 95\% confidence region
obtained by marginalizing over $\Omega_m$ and collapsing to a one-parameter fit
to $w$. (This is due to the fact that $\Delta \chi^2$ criterion for 95\%
confidence depends on the dimensionality of the parameter space.)  

These simple cases demonstrate the differences in the concordance and MLE
procedures and, especially,   some problems which can arise in MLE analysis.
Because we maintain the position that systematic uncertainties dominate the
errors in constructing the constraint regions,  we advocate the concordance
approach as being more conservative for cosmological analysis at the present
time. In general, caution must be exercised when the two methods disagree
significantly, and  the source of the discrepancy must be understood in order
to determine the  true acceptance region.

\eject

\clearpage
\centerline{\null}
\vskip6.2truein
\includegraphics{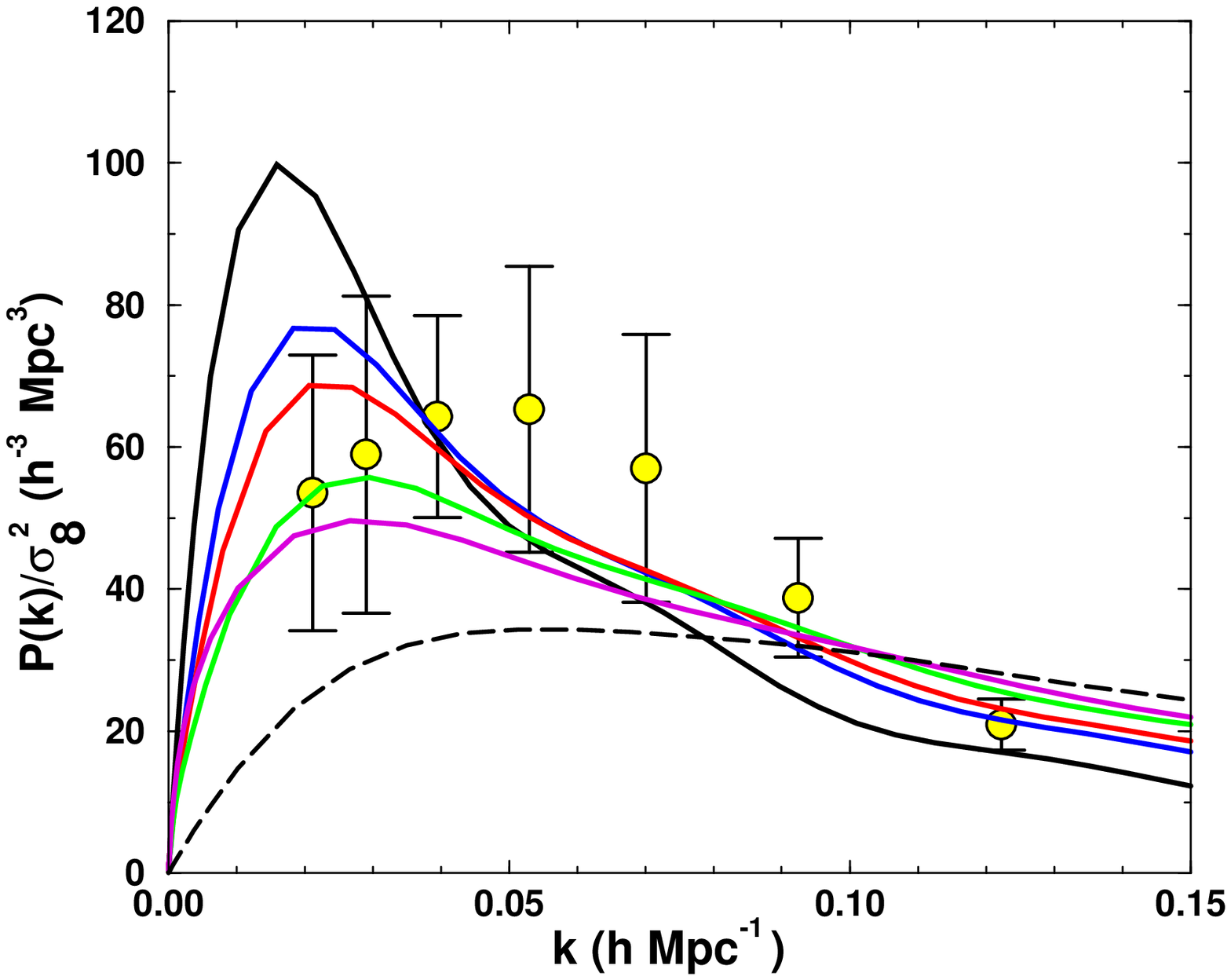}
\figcaption{ The solid circles with $1\sigma$ error bars are the Peacock APM
data we use to test the power spectra shape.  The five solid lines are
quintessence models that pass the shape test with a confidence level of 95\%. 
The model parameters for the black, blue, red, green, and purple curves are: 
$w=$ -1, -1/2, -1/3, -1/6, 0; $\Omega_Q=$ 0.70, 0.60, 0.55, 0.43, 0.20;
$\Gamma=$ 0.20, 0.26, 0.29, 0.37, 0.52. The bias factor is optimized for each
model shown in order to obtain the minimum $\chi^2$: $b=$ 1.01, 1.24, 1.46,
1.81, 2.49, respectively. All models have $n_s=1$ and $\Omega_b h^2 = 0.02$.
For comparison, the dashed line is a standard CDM model with best-fit bias 0.8
which fails the shape test. \label{fig:shape}}
 
\clearpage
\centerline{\null}
\vskip6.2truein
\includegraphics{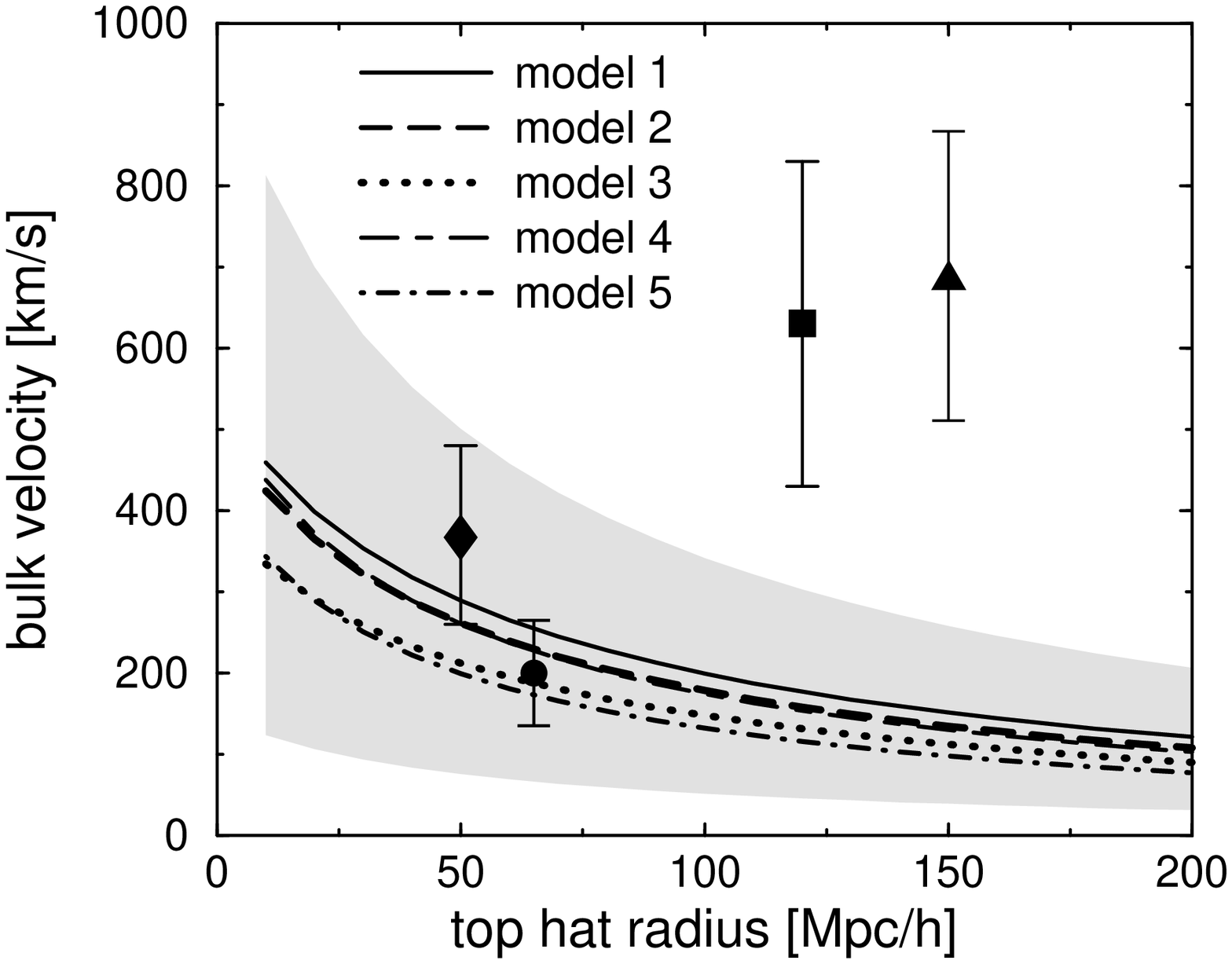}
\figcaption{
The  bulk velocity  predictions as a function of radius (assuming a top hat
window function) are shown for a representative set of QCDM models, and
$\Lambda$CDM  given at the end of this paper in Table~I.  Surrounding the
best-fit quintessence model (Model 2), we have shown the shaded sheath
corresponding to the $95\%$CL, according to a Maxwellian distribution of bulk
velocities. The diamond, circle, square, and  triangle  show the bulk
velocities measured by \cite{Dekel98,Giovanelli98,Hudson99,LP94} with $2
\sigma$ error bars, respectively. We have idealized the window function for the
galaxy velocity catalog, assuming a spherical top hat. \label{fig:bulkv}}

\clearpage
\centerline{\null}
\vskip6.2truein
\includegraphics{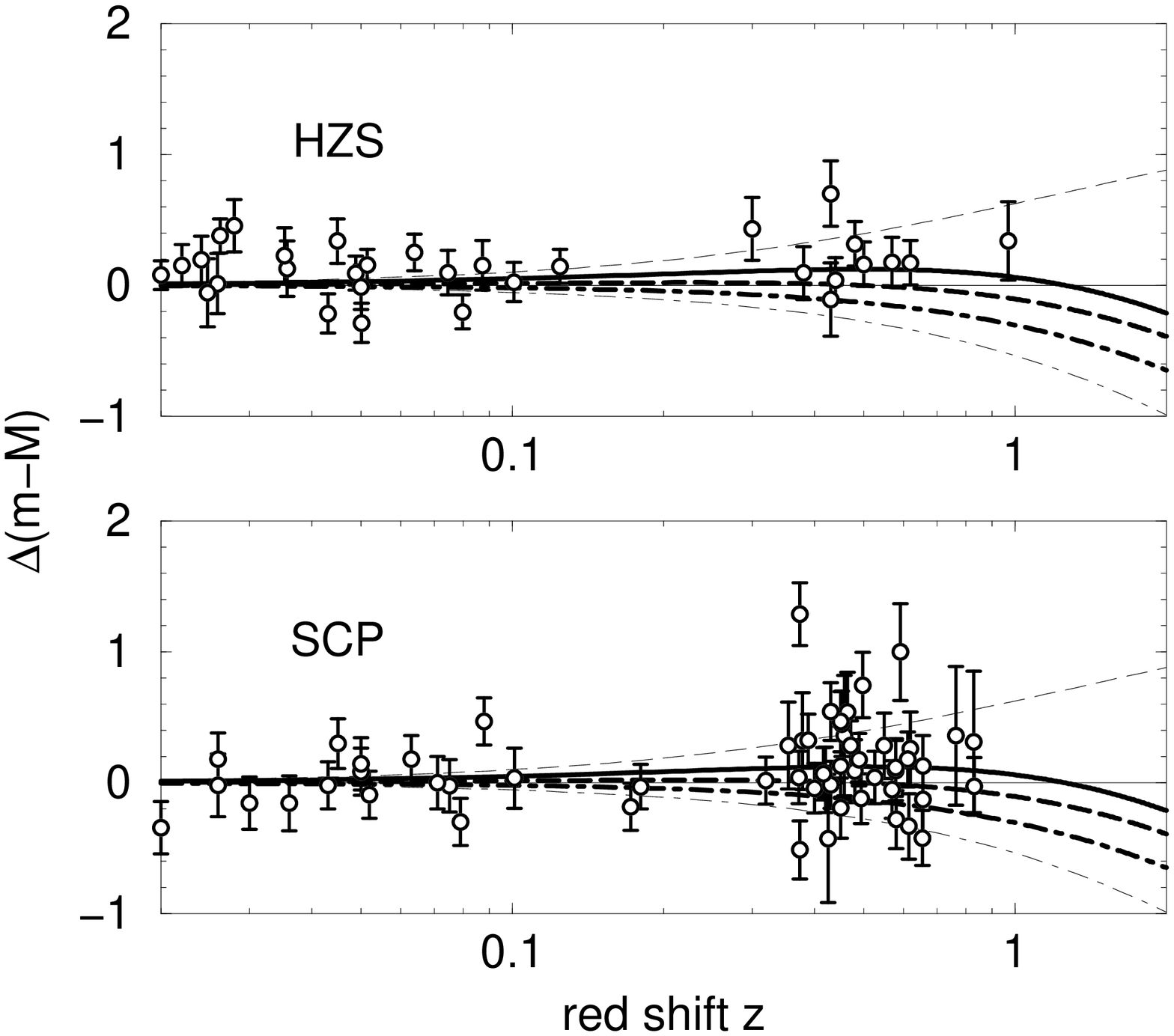}
\figcaption{
The magnitude - red shift relationship determined by type 1a SNe is shown, for
the HZS (using the MLCS analysis method) above, and SCP (full data  set)
below.  The horizontal $\Delta (m-M) = 0$ reference line shows the prediction
of an empty universe ($\Omega_{total}=0$), which has been subtracted from all
data and theoretical curves.  The thin dashed and dot-dashed curves show the
predicted magnitude-red shift relationship for flat models with
$\Omega_\Lambda=1$ and $\Omega_m=1$, respectively.  The vertical offset of the
data has been determined by minimizing the $\chi^2$ to the best fit
$\Lambda$CDM model with $\Omega_m=0.3$, which is given by the thick, solid
curve.  The predictions for quintessence (QCDM) models with $w=-2/3,\, -1/3$
for the same matter density are shown by the thick dashed and dot-dashed
curves.  \label{fig:double_SNe}}

\clearpage
\centerline{\null}
\vskip6.2truein
\includegraphics{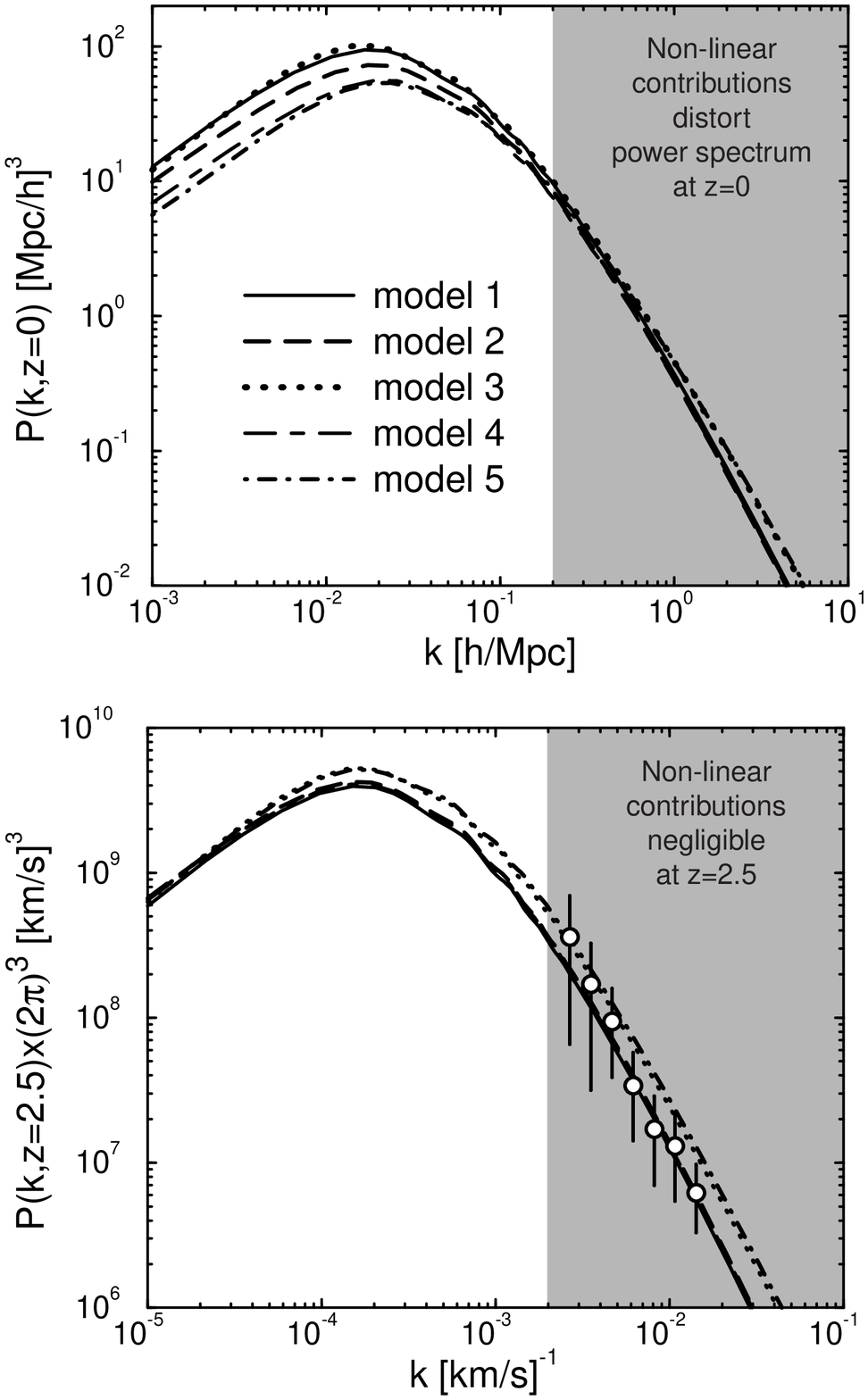}
\figcaption{The upper panel compares the linear mass power spectrum at $z=0$
for the  representative $\Lambda$CDM and QCDM models in Table I.   All models
are COBE normalized and satisfy the cluster abundance constraint on
$\sigma_8$.  The solid and dashed curves have $n_s=1$; the dotted and 
dot-dashed curves have $n_s=1.2$. The shaded region in the top panel indicates
where non-linear contributions are non-negligible. The lower panel shows the
same power spectra projected back in time to  red shift $z=2.5$ and rescaled by
the appropriate value of $h$ at red shift $z$.   Note that, in converting to
$z=2.5$, the abscissa in the lower panel has been rescaled so that it is 
expressed in terms of velocity; once this model-dependent rescaling is made,
the models can be compared directly to the data.   We show the constraints on
the power spectrum, with $1 \sigma$ error bars, as deduced from the Ly-$\alpha$
forest.  Among our representative models, the $n_s=1$ models are preferred over
the $n_s=1.2$ models. \label{fig:lya}}

\clearpage
\centerline{\null}
\vskip6.2truein
\includegraphics{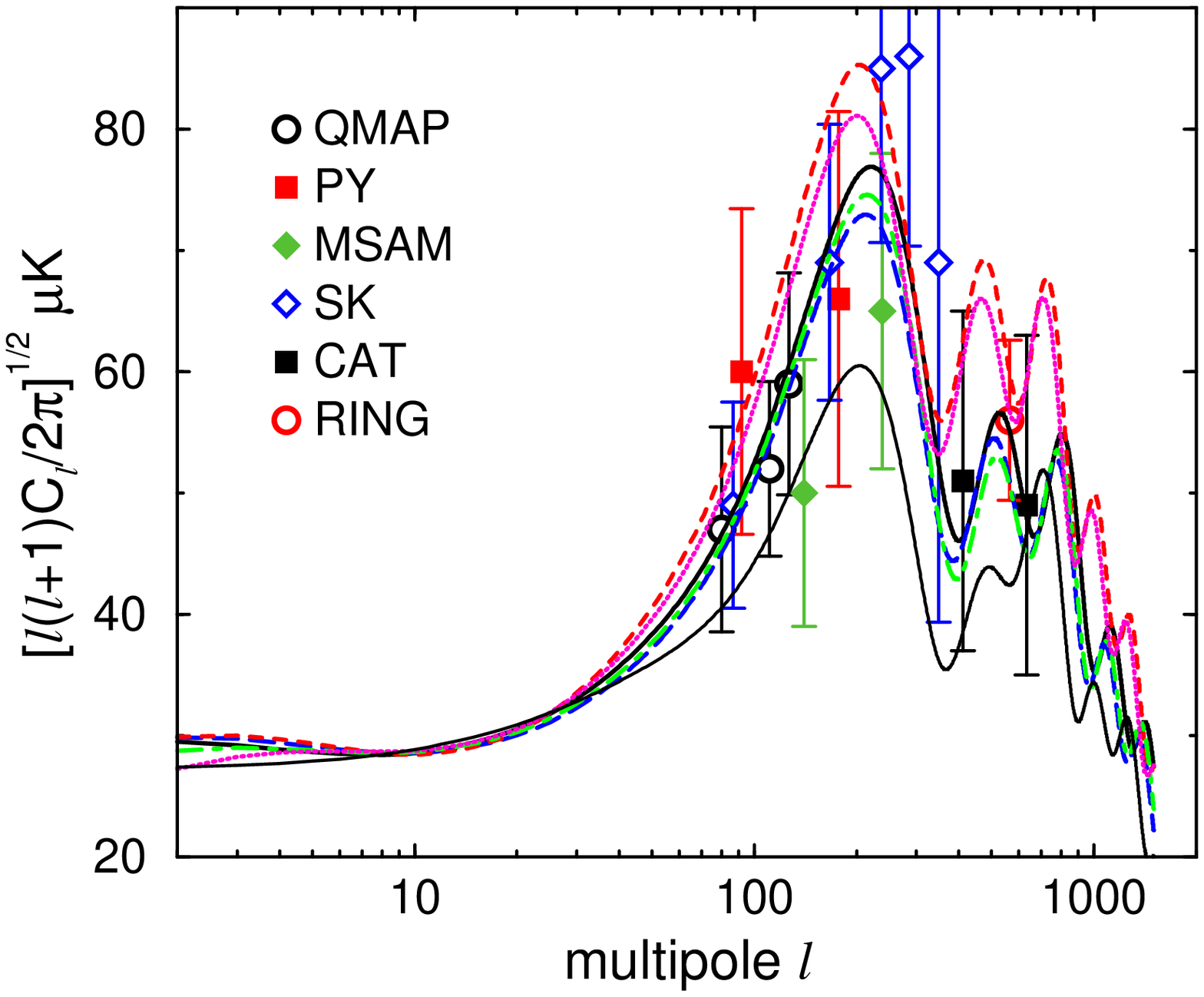}
\figcaption{
The bandpower averages used as the basis of the small angle CMB constraint are
shown.  For comparison, the black, blue, red, green, and purple curves are
models 1-5, given at the end of this paper in Table~I.  The differences are
small, but distinguishable in near-future experiments.  All have acoustic peaks
lying significantly  above those of the standard cold dark matter model (thin
black). The higher $\ell$ results favor our representative models with $n_s=1$
over those with $n_s=1.2$. \label{fig:cmb}}

\clearpage
\centerline{\null}
\vskip6.2truein
\includegraphics{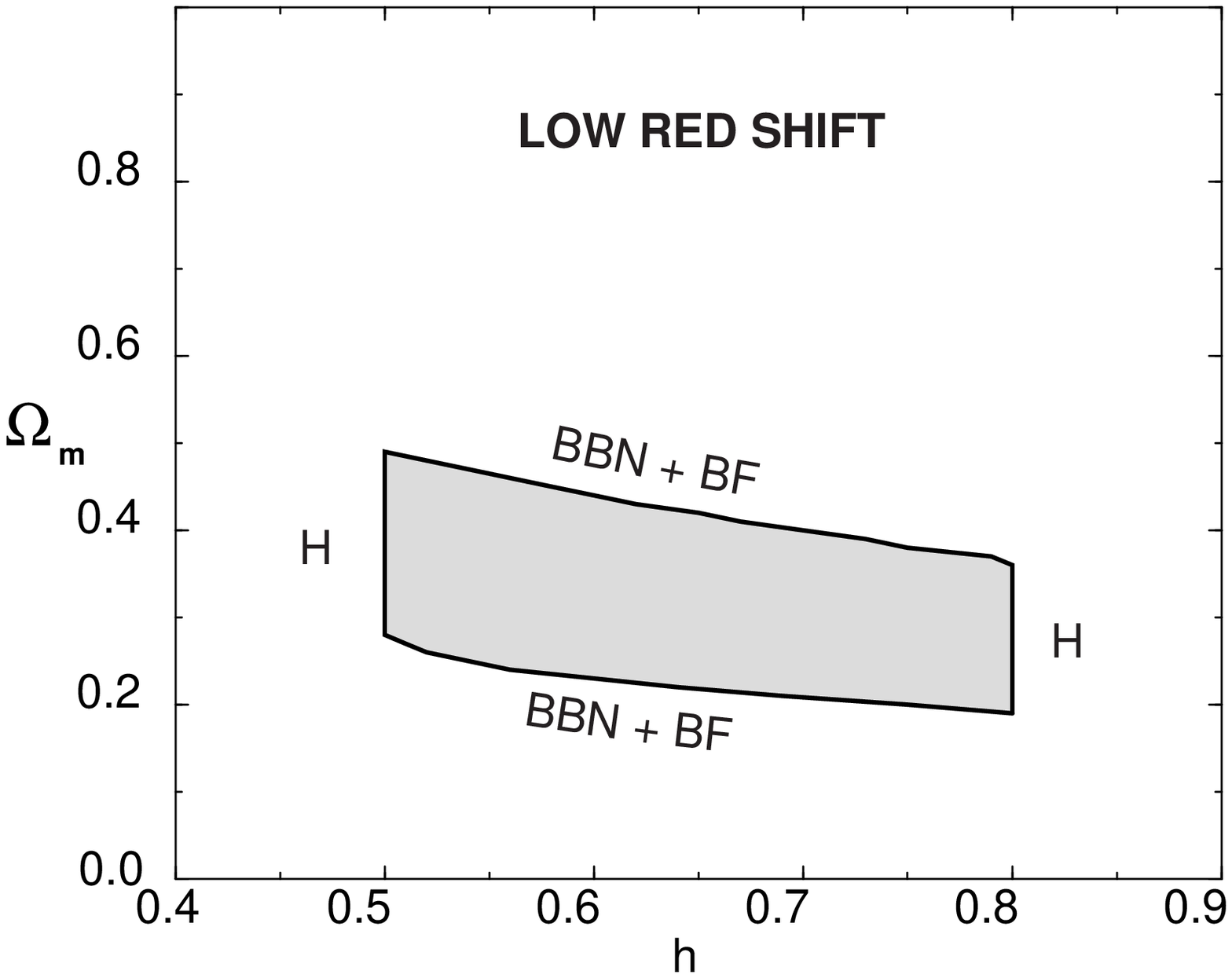}
\figcaption{ The projection of the concordance region on the $\Omega_m - h$
plane, on the basis of the low red shift observational constraints only, is
shown.  The observations which dominate the location of the boundary are
labeled. \label{fig:om-h}}

\clearpage
\centerline{\null}
\vskip6.2truein
\includegraphics{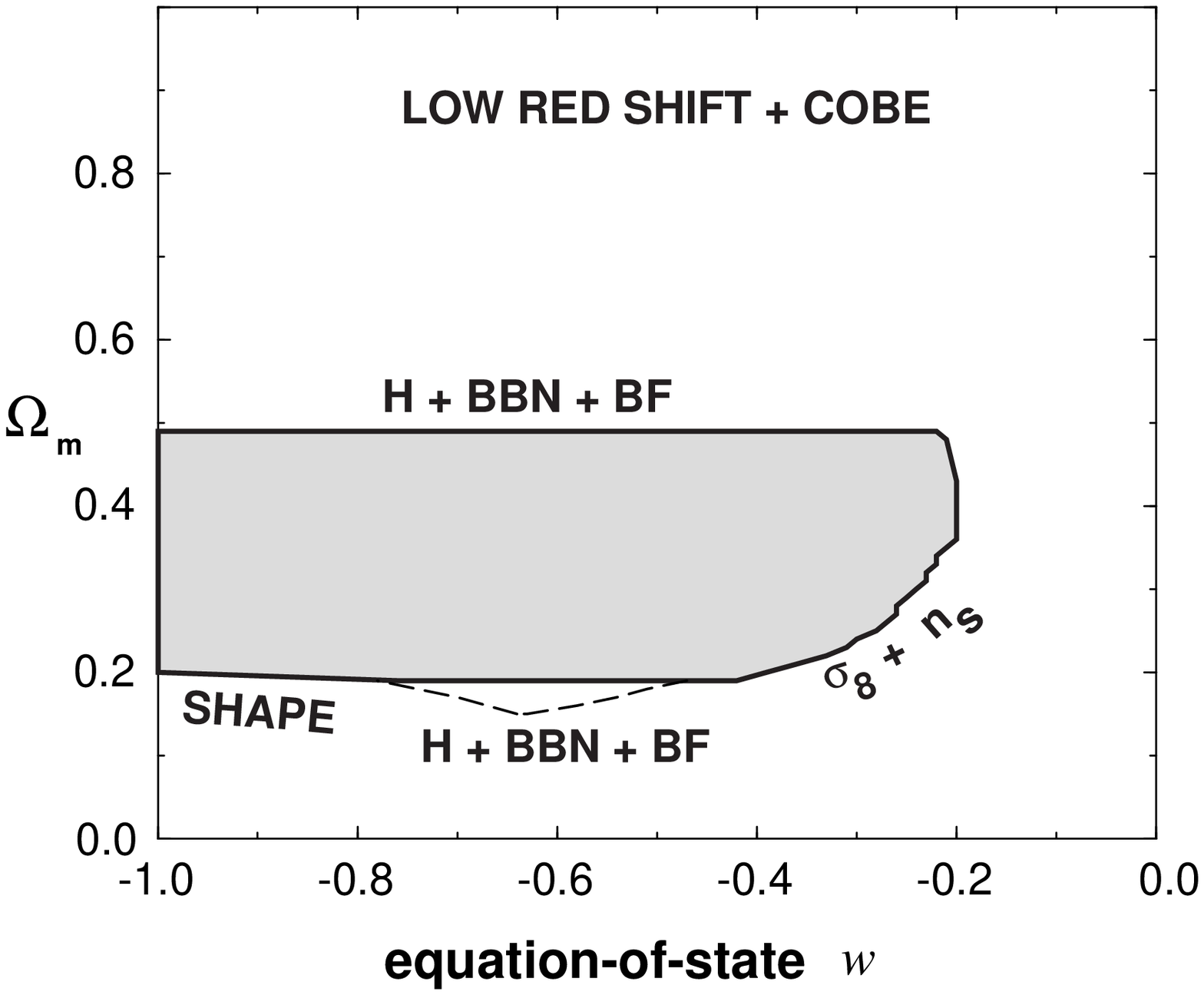}
\figcaption{
The projection of the concordance region on the $\Omega_m - w$ plane, on the
basis of the low red shift and COBE observational constraints only, is shown. 
The observations which dominate the location of the boundary are labelled. If a
wider range for the baryon density is allowed, such as $0.006  < \Omega_b h^2 <
0.022$, the shape test and $\sigma_8$ constraint determine the location of the
low $\Omega_m$ boundary, and the concordance region extends slightly  as shown
by the light dashed line. \label{fig:om-w}}

\clearpage
\centerline{\null}
\vskip6.2truein
\includegraphics{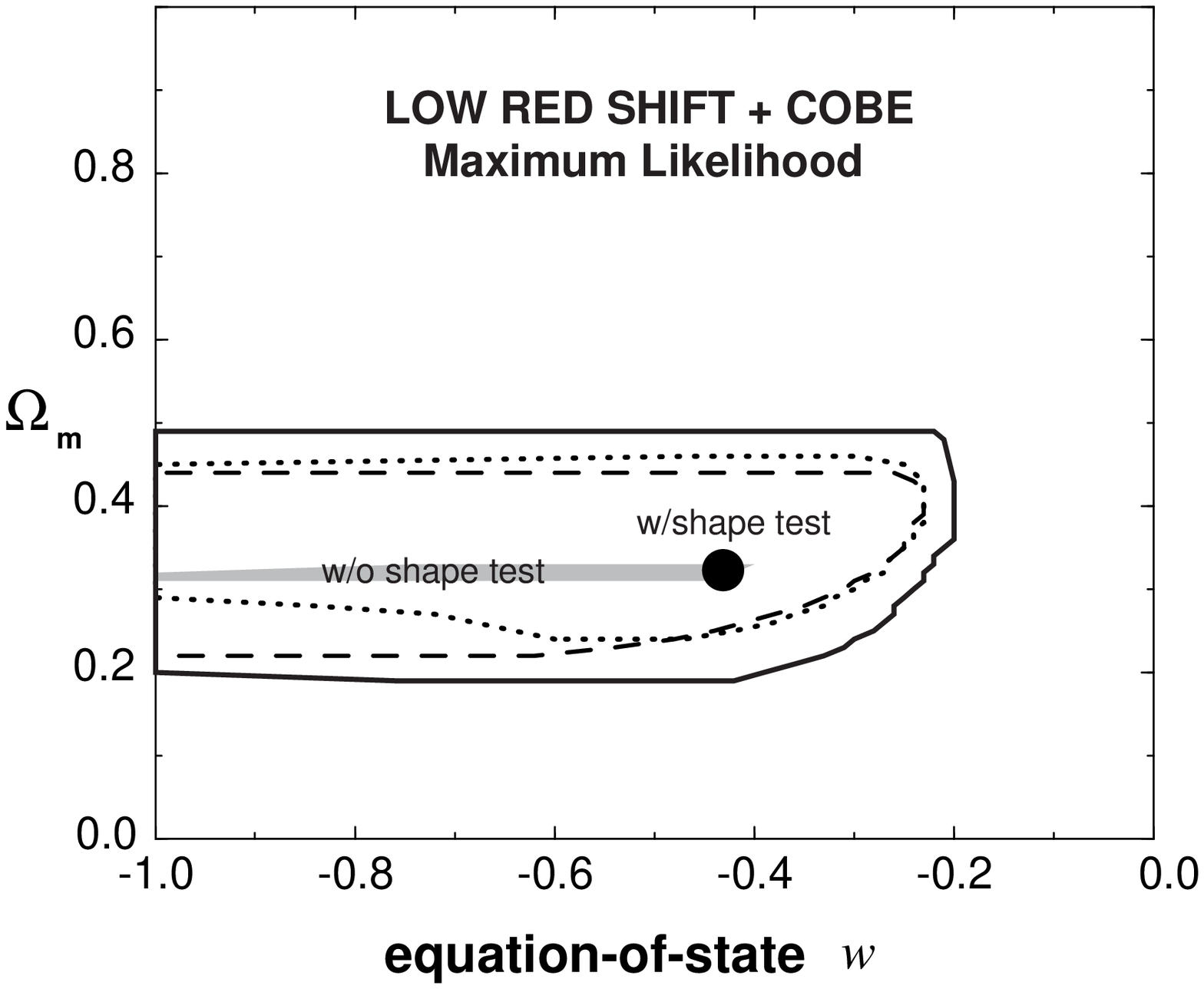}
\figcaption{
The $2\sigma$ maximum likelihood contours in the $\Omega_m - w$ plane with the
low red shift and COBE observational constraints only are shown.  The dotted
and dashed curves show the likelihood contours with and without the mass power
spectrum shape test.  The set of models which maximize the likelihood in each
case are shown by the solid circle and the thick line.   The shape test pushes
models away from $w = -1$, towards $w \sim -1/2$.  The   solid line shows the
$2\sigma$ allowed region according to the concordance region for comparison.
\label{fig:om-w.mle}}

\clearpage
\centerline{\null}
\vskip6.2truein
\includegraphics{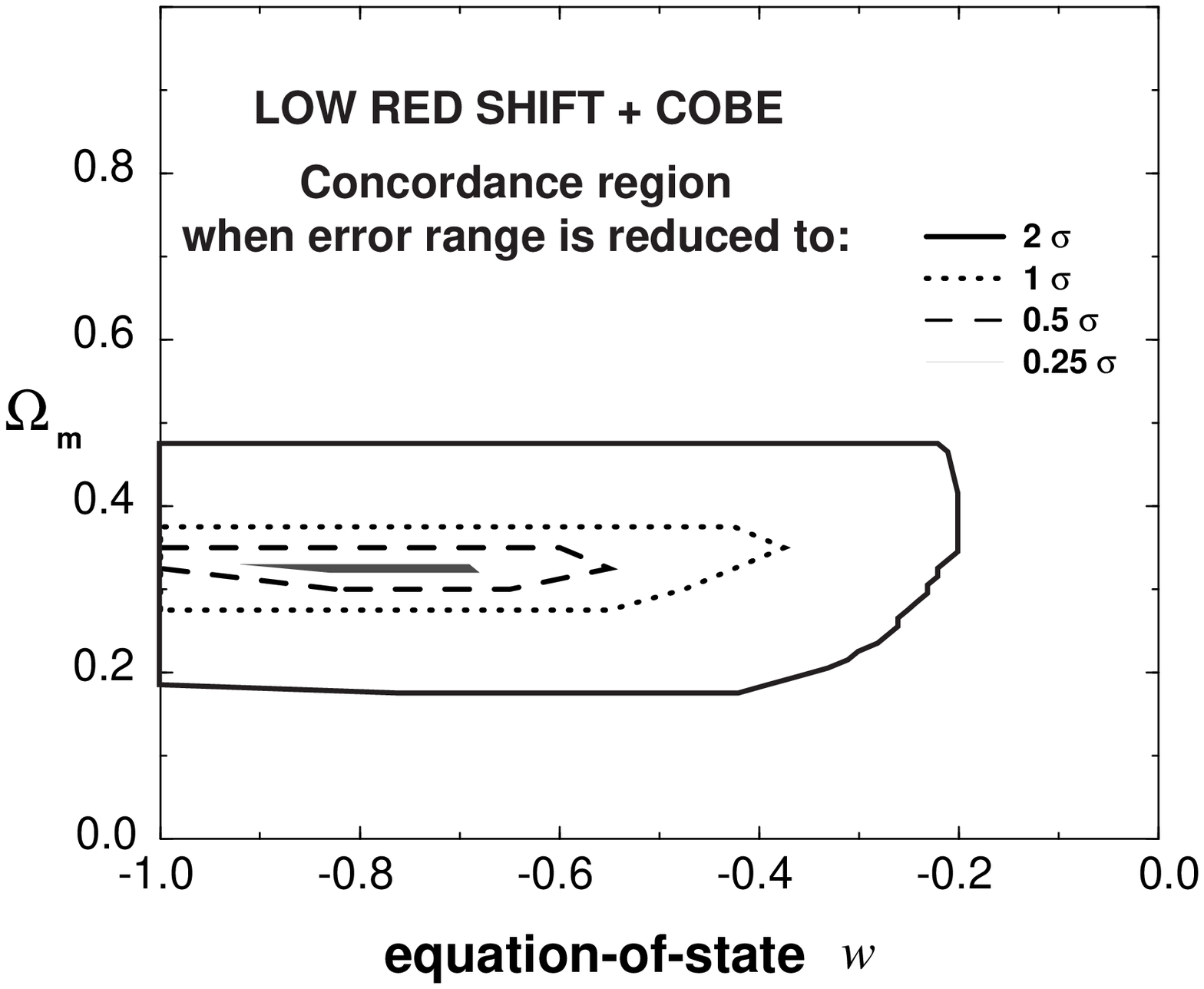}
\figcaption{
We carry out the exercise of shrinking the error bars on all measurements to
obtain the equivalent `best fit' models in the concordance approach.  The
surviving models have $\Omega_m \sim 0.33$ and $-0.9 < w < -0.7$.  This is
similar to the best fit models obtained from the maximum likelihood method,
however the concordance models are not as strongly affected by the shape test.
\label{fig:om-w.shrink}}

\clearpage
\centerline{\null}
\vskip6.2truein
\includegraphics{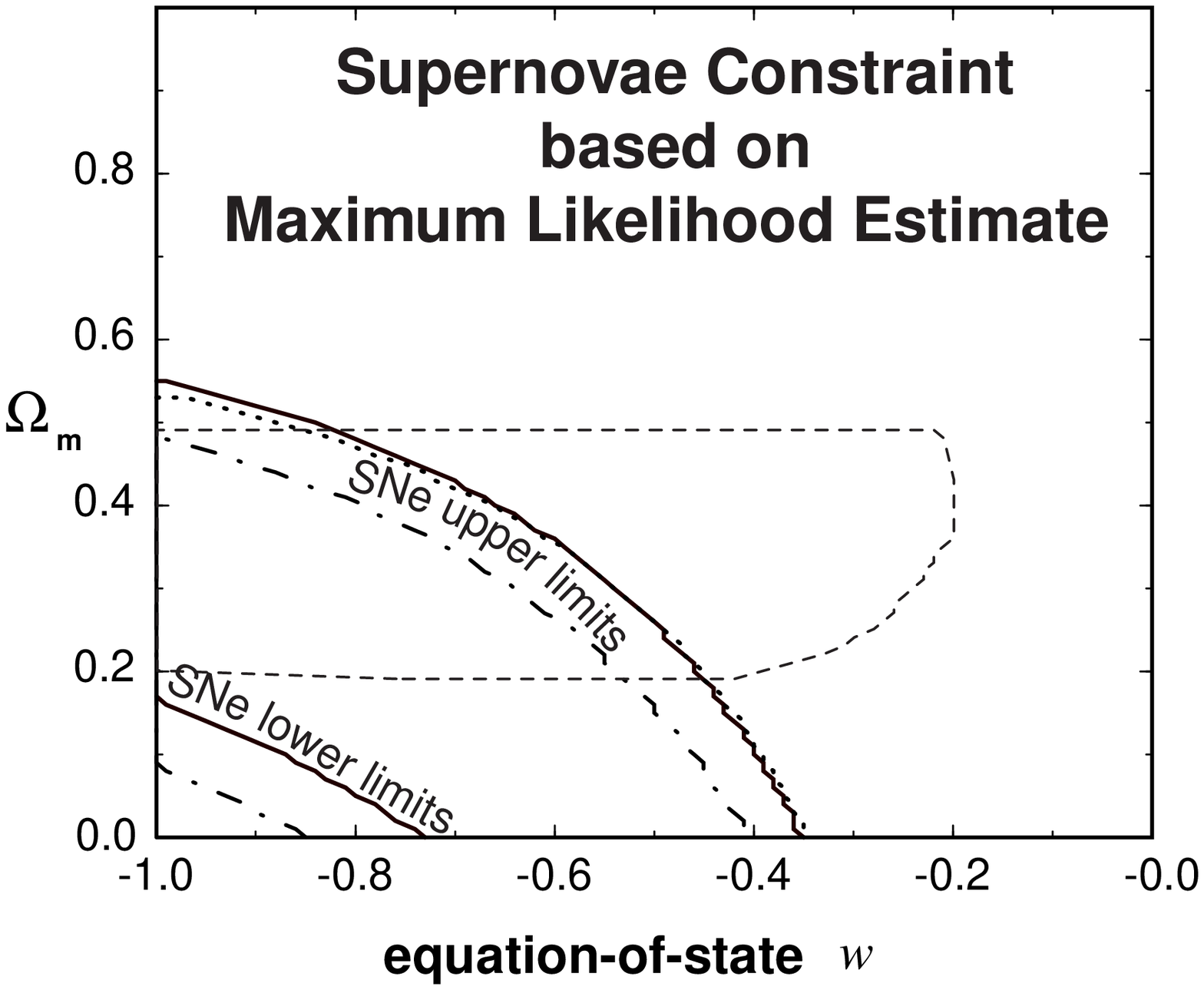}
\figcaption{
The $2\sigma$ maximum likelihood constraints on the $\Omega_m - w$ plane, due
to the SCP (solid), HZS MLCS (short dashed), and HZS template fitting methods
(dot-dashed).  The light, dashed line shows the low red shift concordance
region. \label{fig:om-wsnemle}}

\clearpage
\centerline{\null}
\vskip6.2truein
\includegraphics{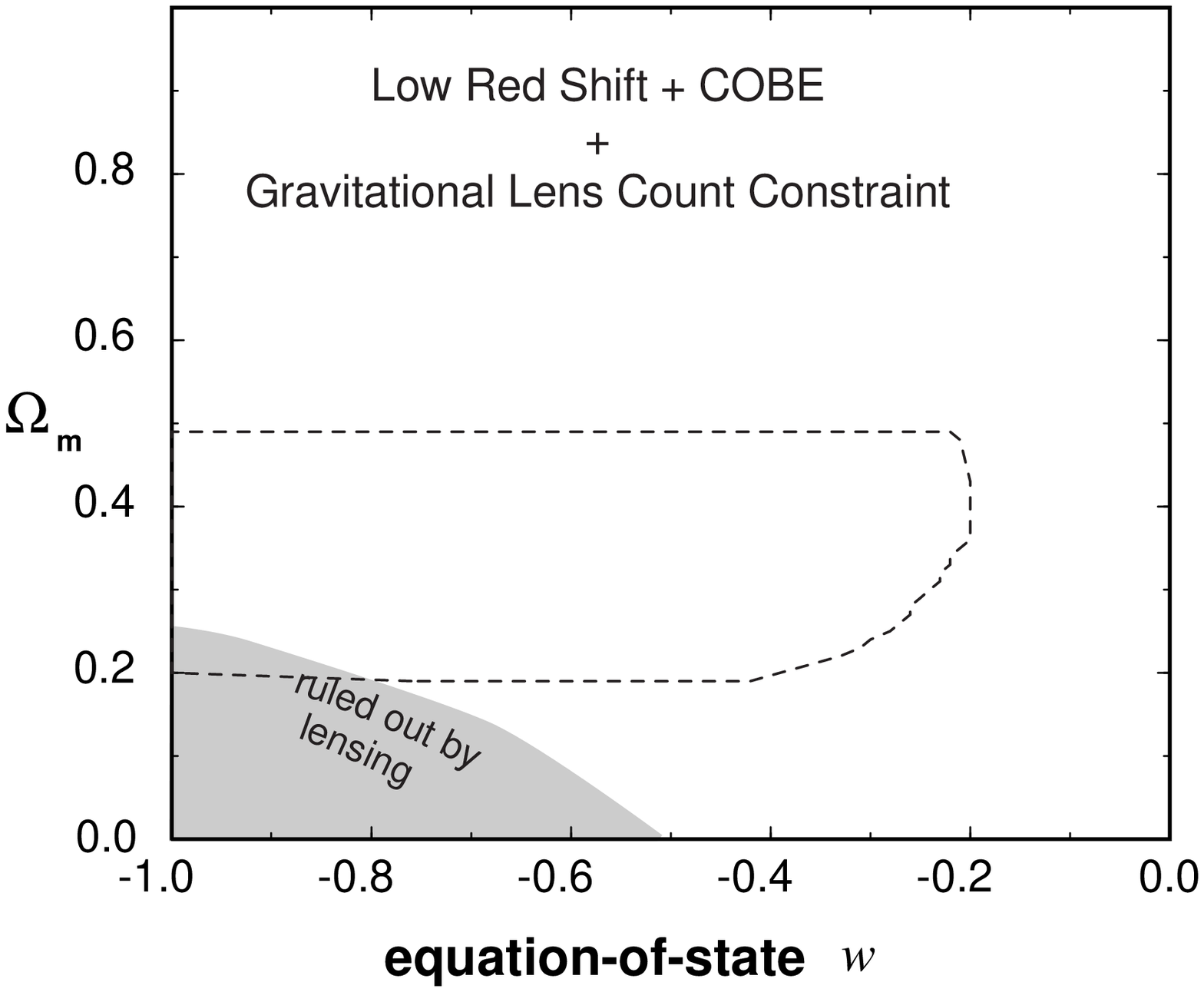}
\figcaption{
The $2\sigma$ gravitational lensing constraint on the low red shift concordance
region is shown.  \label{fig:lensing}}
 
\clearpage
\centerline{\null}
\vskip6.2truein
\includegraphics{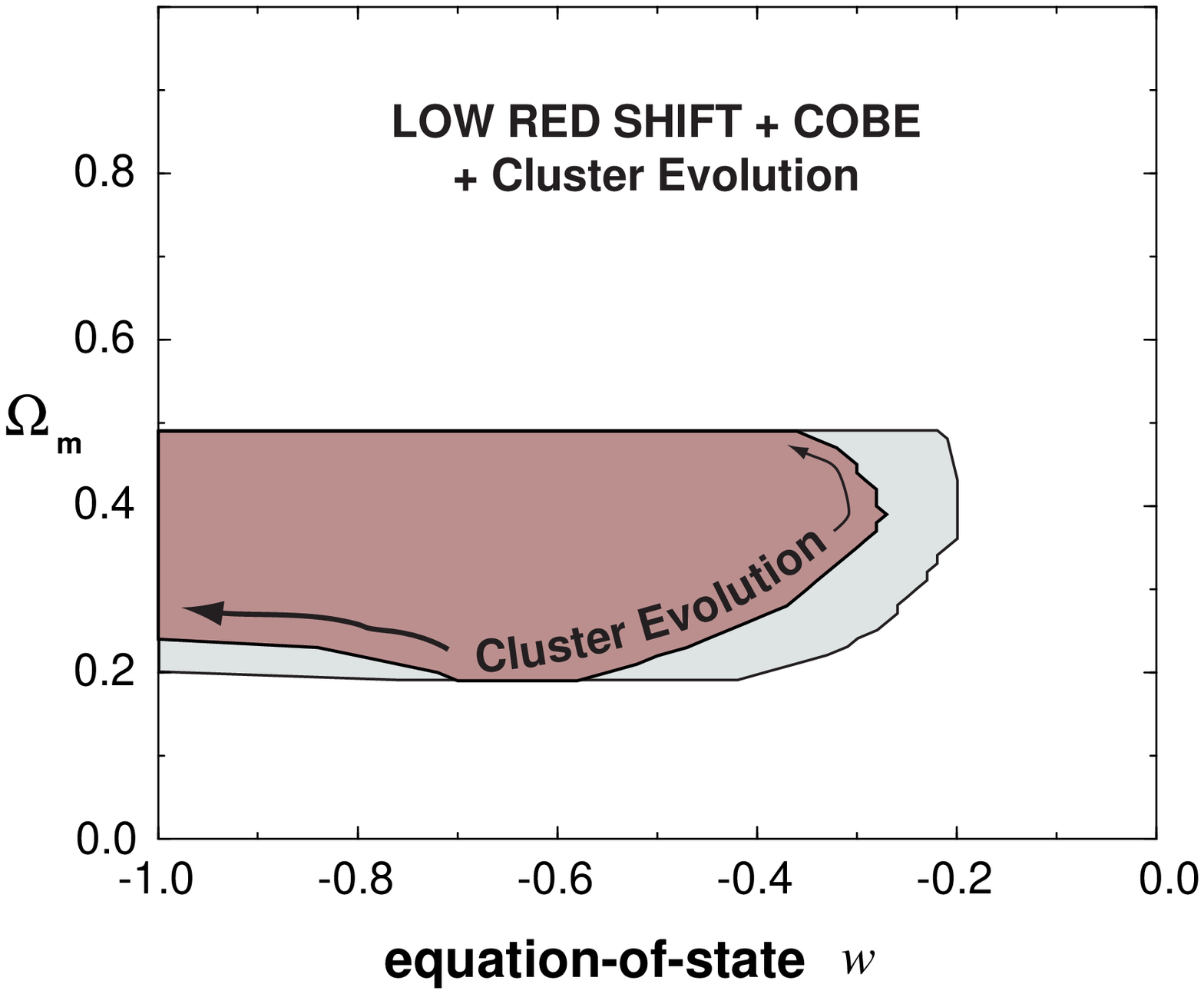}
\figcaption{
The effect of the x-ray cluster abundance evolution constraint on the
projection of the concordance region to the $\Omega_m - w$ plane is shown by
the dark shaded region.  The light shaded region is due to the low red shift
constraints only.   \label{fig:om-w.cluster}}

\clearpage
\centerline{\null}
\vskip6.2truein
\includegraphics{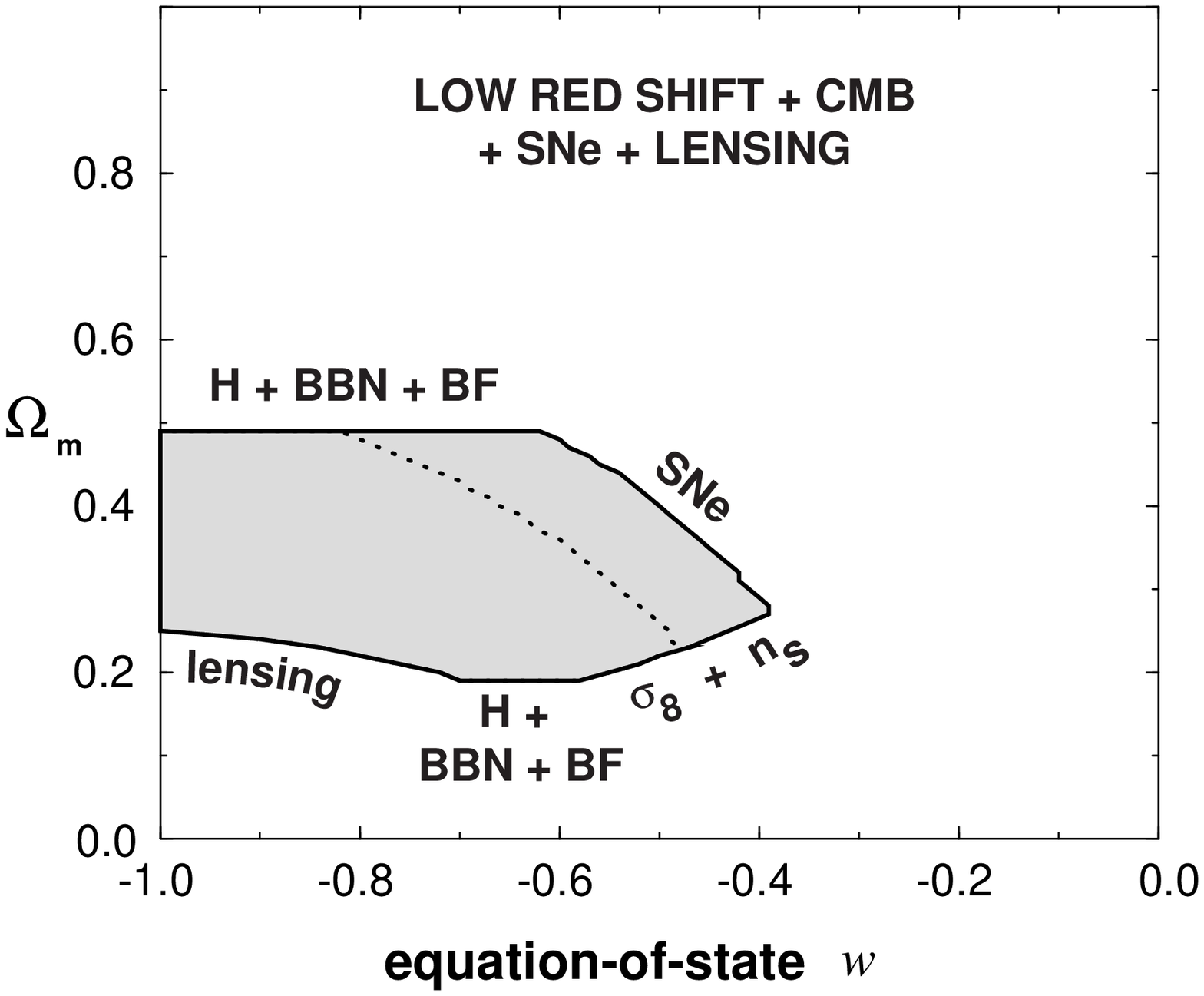}
\figcaption{
The dark shaded region is the projection of the concordance region on the
$\Omega_m - w$ plane with the low, intermediate, and high red shift
observational constraints.  The dashed curve shows the 2$\sigma$ boundary as
evaluated using maximum likelihood, which is the same as Figure 10  (See the
Appendix for a comparison of the tests and a discussion of the pitfalls of the
maximum likelihood approach.) \label{fig:ult_concord}}

\clearpage
\centerline{\null}
\vskip6.2truein
\includegraphics{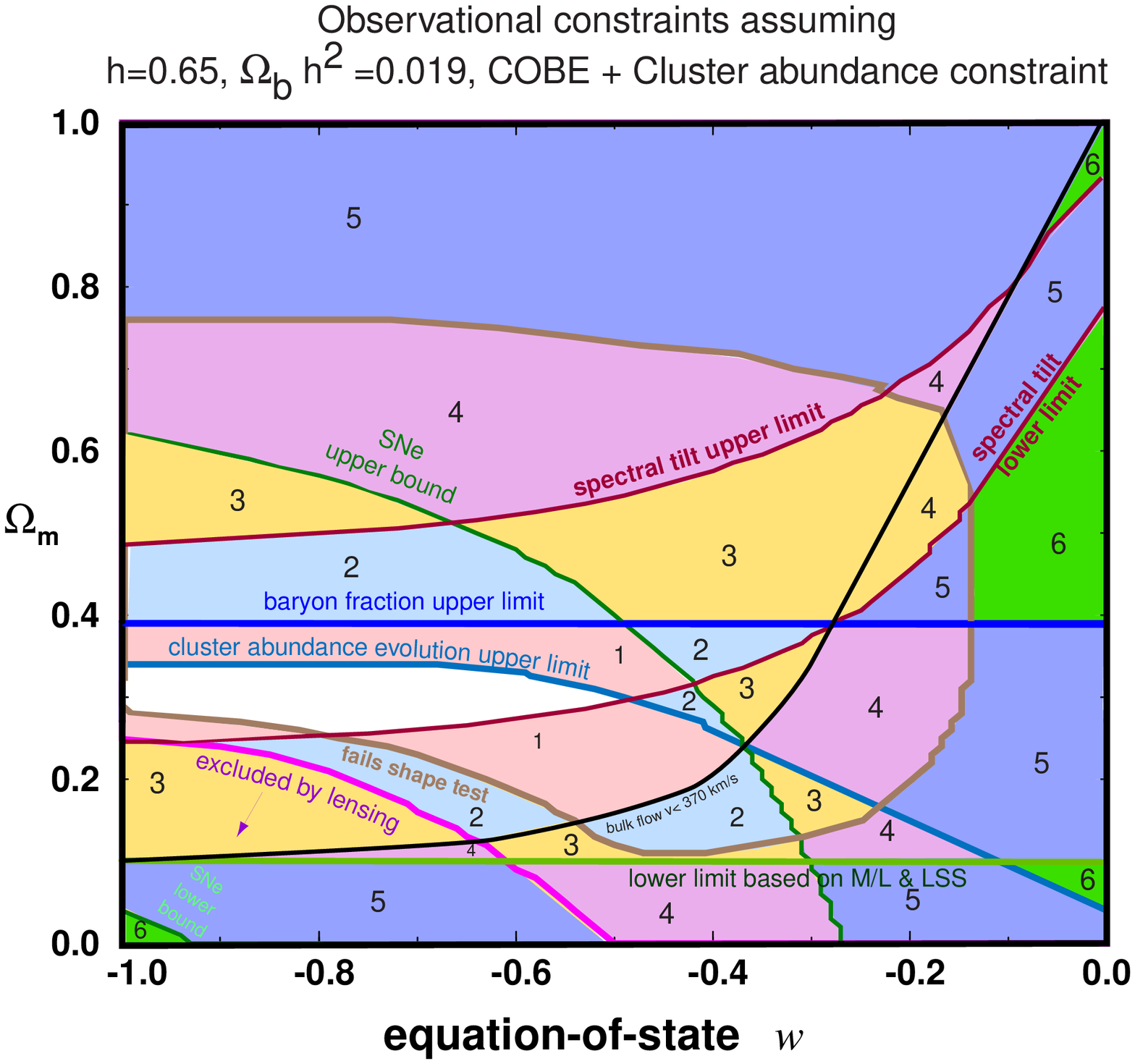}
\figcaption{
The concordance region (white) resulting if we artificially set $h=0.65$ and 
$\Omega_b h^2 =0.019$ precisely and fix the spectral tilt to  precisely match
the central values of COBE normalization and  cluster abundance measurements.
The curves represent the constraints imposed by individual measurements. The
curves divide the plane into patches which have been numbered (and colored)
according to the number of constraints violated by models in that patch. 
\label{fig:picasso}}
 
\clearpage
\centerline{\null}
\vskip6.2truein
\includegraphics{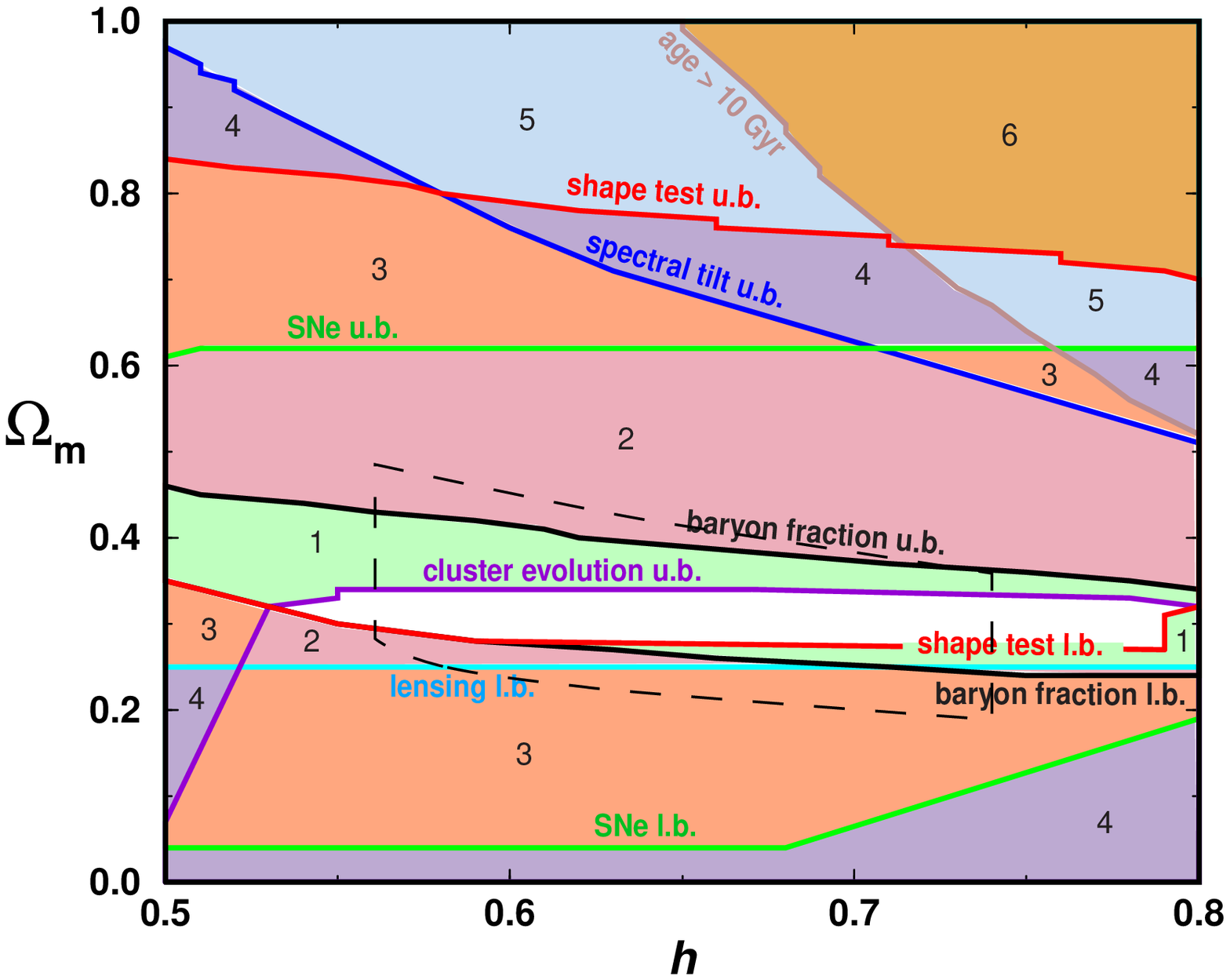}
\figcaption{The concordance region (white) resulting if we artificially set  
$\Omega_b h^2 =0.019$ and fix the spectral tilt to  precisely match
the central values of COBE normalization and  cluster abundance measurements.
The curves represent the constraints imposed by individual measurements. The
curves divide the plane into patches which have been numbered (and colored)
according to the number of constraints violated by models in that patch. 
\label{fig:picasso2}}
 
\clearpage
\clearpage
\centerline{\null}
\vskip6.2truein
\includegraphics{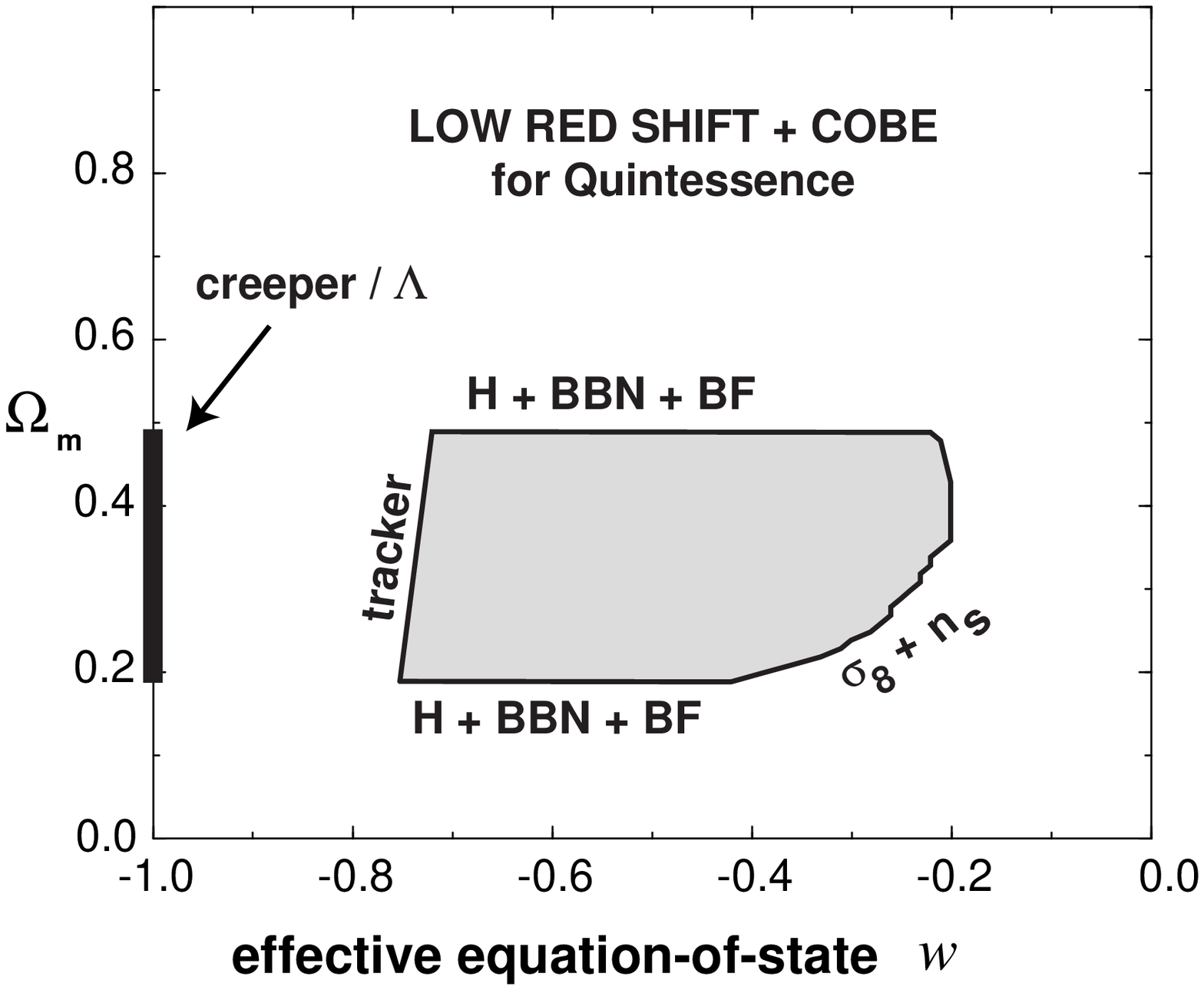}
\figcaption{ The concordance region based on COBE and low red shift tests for
tracker quintessence is shown. The thin black swath along $w=-1$ shows the
allowed region for creeper quintessence and $\Lambda$. The equation-of-state is
time-varying; the abscissa is the effective (average) $w$. \label{fig:tracker}}

\clearpage
\centerline{\null}
\vskip6.2truein
\includegraphics{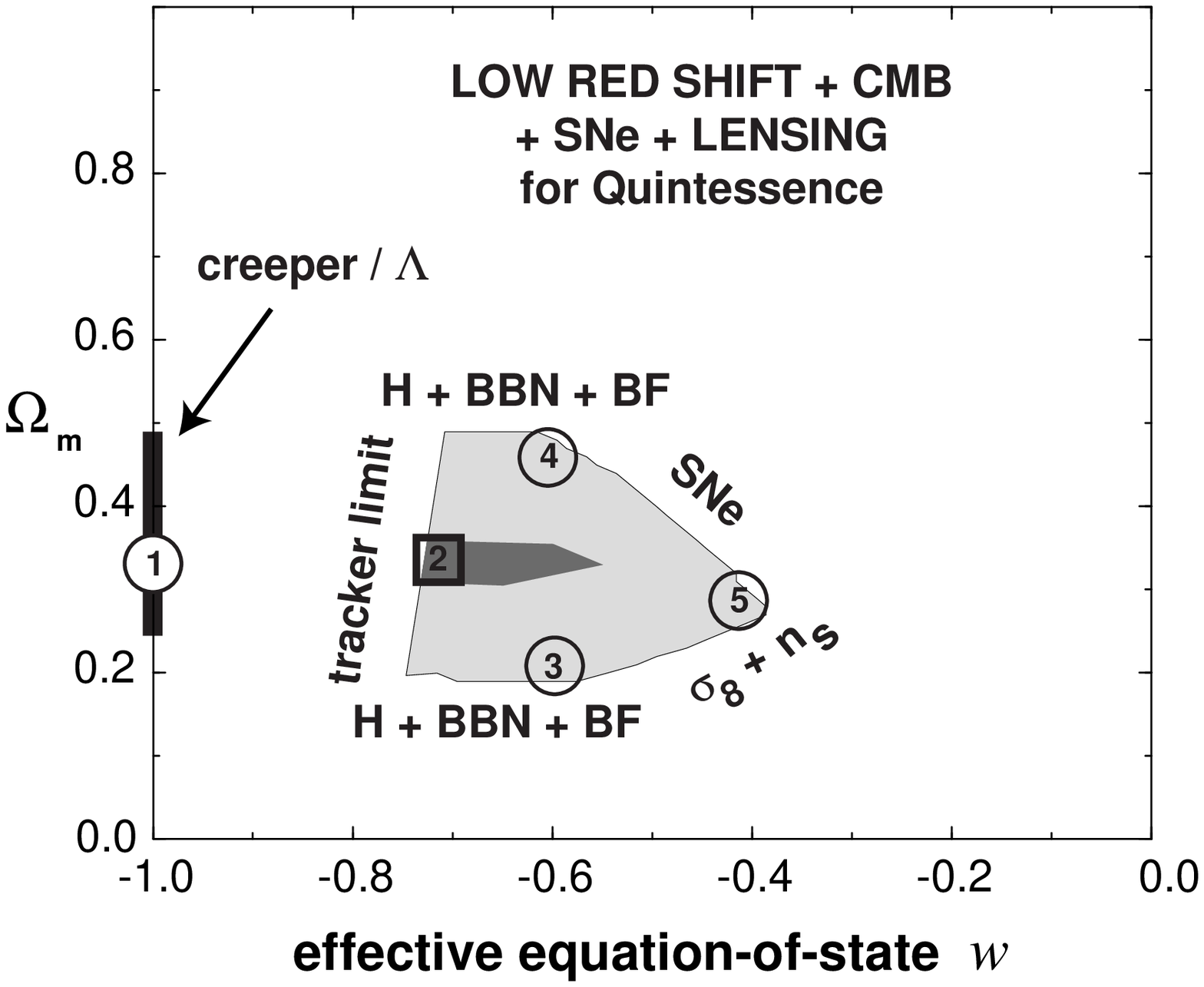}
\figcaption{ The overall concordance region based  low, intermediate, and high
red shift tests for tracker quintessence is shown.  The thin black swath along
$w=-1$ shows the allowed region for creeper quintessence and $\Lambda$. The
equation-of-state is time-varying; the abscissa is the effective (average) $w$
as defined in Eq.~(\protect{\ref{effect}}).  The dark shaded region corresponds
to  the  most preferred  region (the 2$\sigma$ maximum likelihood region
consistent with the tracker constraint),  $\Omega_m \approx 0.33 \pm 0.05$,
effective equation-of-state $w \approx -0.65 \pm 0.10$ and  $h=0.65 \pm 0.10$
and are consistent with spectral index $n=1$. The numbers refer to the
representative models that appear in Table I and that are  referenced
frequently in the text. Model 1 is the best fit $\Lambda$CDM model and Model 2
is the best fit QCDM model.  \label{fig:stracker}}

\clearpage
\centerline{\null}
\vskip6.2truein
\includegraphics{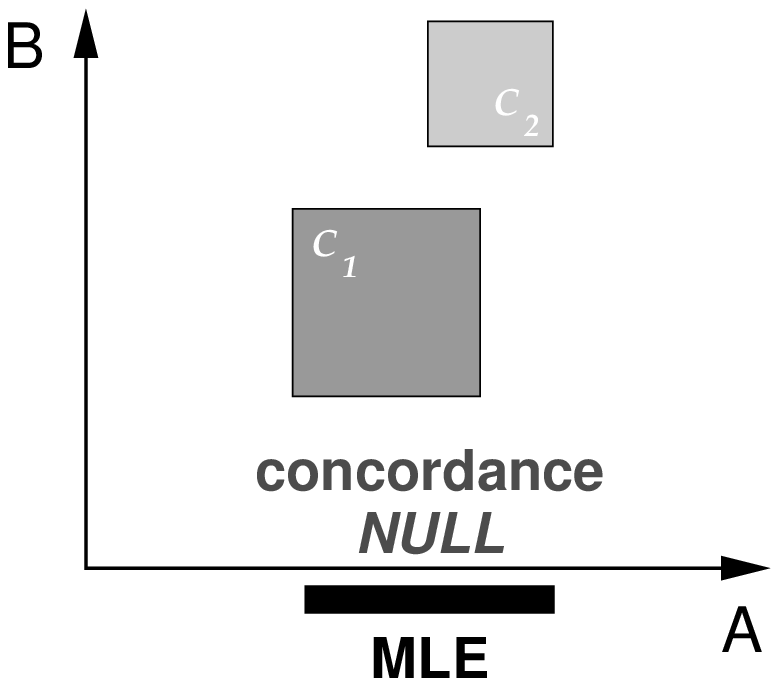}
\figcaption{
The constraint regions $C_1$ and $C_2$ do not intersect. In the concordance
method for determining bounds on $A$, we first find the intersection of the
$C_1$ and $C_2$ in the full, higher-dimensional parameter space, and then we
project that intersection region to obtain the constraint on $A$. In this case,
the concordance region is null. The MLE method always allows some finite region
of $95\%$CL.
\label{fig:case1}}
 
\clearpage
\centerline{\null}
\vskip6.2truein
\includegraphics{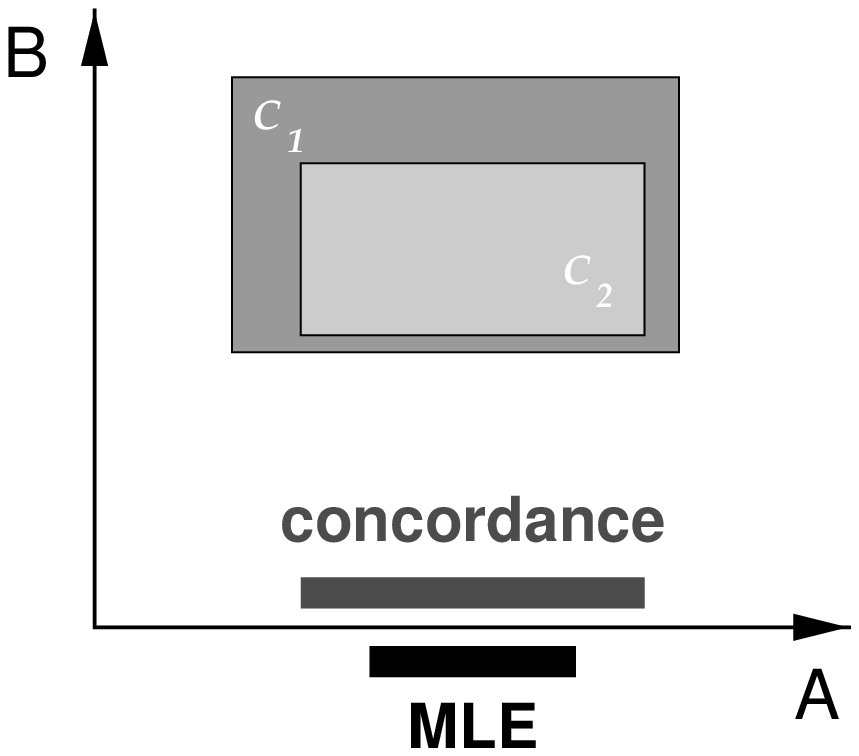}
\figcaption{The constraint regions $C_1$ and $C_2$ overlap. The projection of the
concordance and MLE regions onto parameter $A$, along the horizontal
axis are shown by shaded strips.\label{fig:case3}}

\eject
\clearpage

\begin{figure}
\epsfxsize=7.75in 
\epsfbox{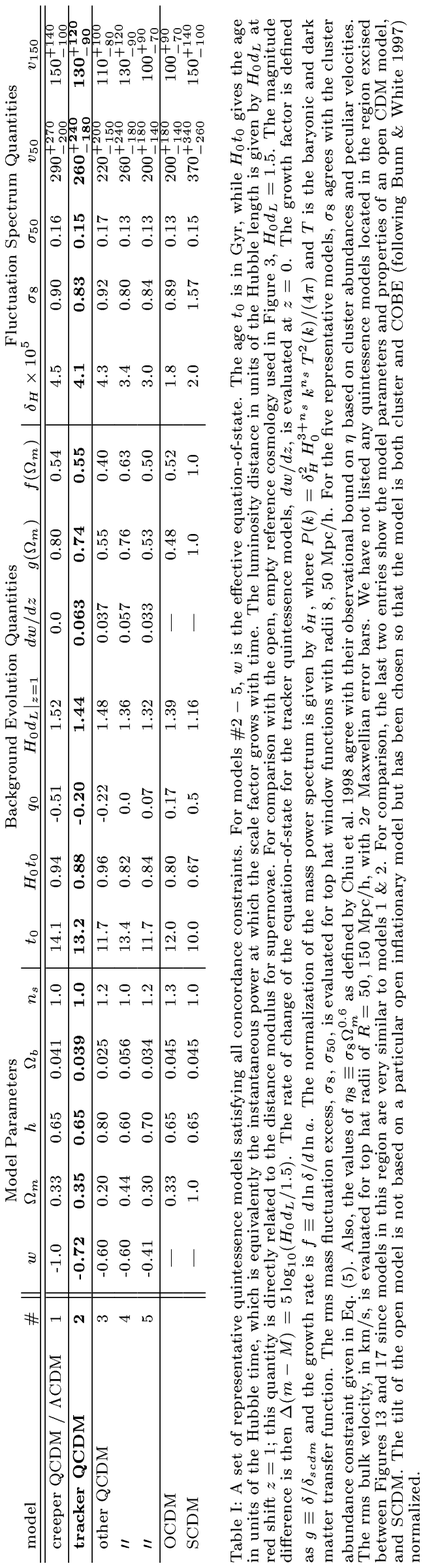}
\caption{\label{tablep}}
\end{figure}

\end{document}